\documentclass[aps,prc,twocolumn,showpacs,article]{revtex4}

\usepackage{amssymb}

\usepackage{amsmath,amssymb,mathrsfs}
\usepackage{graphicx}
\usepackage{amsthm}
\usepackage{amscd}
\usepackage{bm}
\usepackage{dsfont}

\usepackage{appendix}

\usepackage[T2A]{fontenc}
\usepackage[cp1251]{inputenc}

\newcommand{\WI}[2]{#1_{\mathrm{#2}}}
\newcommand{\Lamb}[1]{\WI{\Lambda}{#1}}
\newcommand{\etak}[1]{\WI{\eta}{#1}}
\newcommand{\thetak}[1]{\WI{\theta}{#1}}

\begin{document}

\title{Hydrostatic equilibrium of stars \\ without electroneutrality constraint}

\author{M.~I.~Krivoruchenko,$^{1,2}$ D.~K.~Nadyozhin,$^{1,3}$ A.~V.~Yudin,$^{1,a}$}

\affiliation{
$^{1}$Institute for Theoretical and Experimental Physics$\mathrm{,}$ B. Cheremushkinskaya 25 \\ 117218 Moscow, Russia\\
$^{2}$Bogoliubov Laboratory of Theoretical Physics$\mathrm{,}$ Joint Institute for Nuclear Research \\ 141980 Dubna, Russia \\
$^{3}$Kurchatov Institute$\mathrm{,}$ Akademika Kurchatova Pl.~1\\ 123182 Moscow, Russia \\
$^{a}${E-mail: yudin@itep.ru}}

\begin{abstract}
The general solution of hydrostatic equilibrium equations
for a two-component fluid of ions and electrons without a local electroneutrality constraint is found in the framework of Newtonian gravity theory.
In agreement with the Poincar\'e theorem on analyticity and
in the context of Dyson's argument, the general solution is demonstrated to possess a fixed (essential) singularity in the gravitational constant $G$ at $ G = 0 $.
The regular component of the general solution can be determined by perturbation theory in $G$ starting from a locally neutral solution.
The non-perturbative component obtained using the method of Wentzel, Kramers and Brillouin
is exponentially small in the inner layers of the star and grows rapidly in the outward direction.
Near the surface of the star, both components are comparable in magnitude, and their non-linear interplay
determines the properties of an electro- or ionosphere.
The stellar charge
varies within the limits of $- 0.1 $ to $150$ C per solar mass.
The properties of electro- and ionospheres are exponentially sensitive to variations of the fluid densities in the central regions of the star.
The general solutions of two exactly solvable stellar models without a local electroneutrality constraint are also presented.
\end{abstract}

\pacs{
97.10.-q,
97.10.Cv
}

\maketitle

%%%%%%%%%%%%%%%%%%%%%%%%%%%%%%%%%%%%%%%%%%%%%%%%%%%%%%%%%%%%%%%%%

%***************************************************************
\renewcommand{\baselinestretch}{1.0}
\renewcommand{\theequation}{\thesection.\arabic{equation}}

%%%%%%%%%%%%%%%%%%%%%%%%%%%%%%%%%%%%%%%%%%%%%%%%%%%%%%%%%%%%%%%%%
\section{Introduction}
%%%%%%%%%%%%%%%%%%%%%%%%%%%%%%%%%%%%%%%%%%%%%%%%%%%%%%%%%%%%%%%%%
\setcounter{equation}{0}

Charged particles constitute an appreciable or dominant fraction of the matter in stars.
The electromagnetic interaction between particles is stronger than the gravitational
interaction by a factor of
\begin{equation}
\WI{\lambda}{G} = \frac{e^2}{G m_{\mathrm{u}}^2} \approx 1.25 \times 10^{36}, \label{lambda G}
\end{equation}
where $ e $ denotes the proton charge in CGS units, $ G $ is the gravitational constant,
and $ m_{\mathrm{u}} = 931.5 $ MeV/$c^2 $ is the atomic mass unit, with $ c $ being the speed of light.
Under the assumption of the complete screening of electrostatic interactions, gravitational forces should play the dominant role
in determining stellar structure.
The hydrostatic equilibrium (HE) equations
are usually supplemented with a local electroneutrality constraint (LEC);
however, it is not immediately clear how effective the screening is at compensating for the
enormous difference in the strengths of the electromagnetic and gravitational forces.

Lebowitz and Lieb \cite{Lebowitz1969} showed that the LEC is required %one of the sufficient conditions
for the existence of extensive thermodynamic potentials of isolated systems.
The stability of electrically neutral, isolated systems
of charged particles in the thermodynamic limit was proven by Dyson and Lenard \cite{Dyson1967}
and by Lieb and Thirring \cite{Lieb1975}.

The thermodynamic equilibrium
in an external potential field is determined by Gibbs' condition \cite{Landau1976}:
\begin{equation} \label{Gibbs}
\mu + V(x) = \mathrm {const},
\end{equation}
where $ \mu $ is the chemical potential and $ V (x) $ is the external potential.
A generalization of Eq.~(\ref{Gibbs}) for a multicomponent fluid
within the framework of General Relativity Theory (GRT) was given by O.~Klein \cite {Klein1949}
and, in a broader context, by Kodama and Yamada \cite{Kodama1972} and Olson and Bailyn \cite{Olson1975}.
In the presence of gravity or other external fields, the chemical potentials of a multicomponent fluid acquire coordinate dependence,
as already seen from Gibbs' condition, which is generally incompatible with the LEC.

As early as 1924, Rosseland \cite{Rosseland1924} showed that in Newtonian gravity theory (NGT),
the unconstrained HE equations for an ionized self-gravitating gas
with a constant temperature admit solutions with similar distributions of ions and electrons.
The charge density of the gas was found to be positive and $\WI{\lambda}{G} $ times smaller than the particle number density.
In the context of white dwarfs, the polarization of stars was discussed by Schatzman \cite{Schatzman1958}.

The smallness of the uncompensated charge density justifies the use of the LEC
when describing effects not related to the electrostatic properties of stars.

Based on the variational energy principle of GRT, Olson and Bailyn \cite{Olson1975}
derived a complete system of unconstrained HE equations for a charged multicomponent fluid.
In the stellar structure models discussed by Olson and Bailyn \cite{Olson1978}, Bally and Harrison \cite{Bally1978}, and Neslu\v san \cite{Neslusan2001},
the polarization effect leads to a stellar charge of
$\WI{Q}{s} \sim e \WI{M}{s}/(m_{\mathrm{u}}\WI{\lambda}{G}) \sim 100~\WI{M}{s}/M_{\odot}$ C.

Rotondo et al. \cite{Rotondo2011} and Belvedere et al. \cite {Belvedere2012,Belvedere2014}
showed recently that the components of a fluid have different spatial distributions, leading to the appearance of an electrosphere,
the charge of which is sufficiently large to completely compensate for the positive charge of
the star.

The deviations from local electroneutrality and, accordingly, the appearance of
a large-scale electrostatic field in a star
affects the diffusion and separation of chemical elements in the Sun \cite{Gorshkov2008} and in neutron stars \cite{Beznogov2013}.
The uncompensated stellar charge could influence possible mechanisms of supernova explosions \cite{Nadyozhin2005,Krivoruchenko2011,Nadyozhin2016}.

Electrically charged stars \cite{Rosseland1924} -- \cite{Belvedere2014} are described by particular solutions of the unconstrained HE equations.
Some features of the general solution can be foreseen
on the basis of simple qualitative considerations, as discussed below.

Let us consider a neutron star composed of neutrons, electrons and protons.
Suppose that it is electrically neutral and in chemical equilibrium with respect to the weak interaction.
The gravitational field of the star creates a potential well,
in which a number of electrons $ \Delta\WI{N}{e} $ can be confined,
given that that the energy of their electrostatic repulsion $ \sim e^2 \Delta\WI{N}{e}^2 /\WI{R}{s} $
does not exceed the energy of their gravitational attraction to the star
$ \sim G \WI{M}{s} \Delta \WI{N}{e} \WI{m}{e} / \WI{R}{s} $, where $ \WI{R}{s} $ and $ \WI{M}{s} $ are the radius and mass of the star, respectively, and
$ \WI{m}{e} $ is the electron mass.
This condition implies that
$ \Delta \WI{N}{e} \lesssim G \WI{M}{s} \WI{m}{e} / e^2 \sim 5 \times 10^{17} \WI{M}{s}/M_{\odot}$.
The charge of a neutron star may vary from $ \sim -0.1 $ C to zero.
An excess of electrons, however, would violate the assumption of chemical equilibrium.
The relaxation of the neutron star to a new state through the inverse $ \beta $ decay
is accompanied by the emission of
energy in the form of neutrinos.
The star then turns into a lower-energy state with the same electric charge.
If gravitational stability is not lost during the relaxation process, then the above estimate remains valid.
Considering that the number of electrons $ \Delta \WI{N}{e} $ is many orders of magnitude smaller than
the number of neutrons $ M_{\odot}/m_{\mathrm{u}} \approx 1.2\times 10^{57} $,
the neutron star remains gravitationally stable.
Similar arguments lead to the conclusion that the neutron star is capable
of confining $ \Delta \WI{N}{p} \lesssim G \WI{M}{s} \WI{m}{p} / e^2 \sim 10^{21} \WI{M}{s}/M_{\odot}$ additional protons
with a total charge not exceeding $ \WI{Q}{s} \sim 150~\WI{M}{s}/M_{\odot}$ C.
The electric charge of a neutron star can vary from $ -0.1$ to $150 $ C per solar mass.
Similar considerations apply to white dwarfs.

The general solution of the unconstrained HE equations is expected to describe
gravitationally stable configurations
with a stellar charge of $ -0.1$ to $150 $ C per solar mass.

A single-component self-gravitating fluid of hypothetical
particles of mass $m_{f}$ and charge $ef$, with $f \ll 1$,
is discussed by S.~Ray et al.~\cite{Ray2003}.
A comparison of the gravitational force $\sim G m_{f} \WI{M}{s} $
and the electrostatic repulsion force $\sim (ef)^2 \WI{M}{s}/m_{f}$ acting on an elementary volume of the fluid
yields $f \sim m_{f}/(m_{\mathrm{u}} \sqrt{\WI{\lambda}{G}})$ and a maximum stellar charge of
$ \WI{Q}{s} \sim ef\WI{M}{s}/m_{f} \sim  10^{20} \WI{M}{s}/M_{\odot}$ C.
In nature, stars are composed of multicomponent fluids and are subject to a $\sqrt{\WI{\lambda}{G}}$ times more
stringent constraint on $\WI{Q}{s}$ because of charge separation in the
stellar electric fields.

Another fundamental property of the general solution
follows from the Poincar\'e theorem on analyticity (see, e.g., Ref.~\cite{Ilyashenko2008}) and Dyson's argument \cite{Dyson1952}.

The Poincar\'e theorem on analyticity is an intuitively clear theorem that in rough terms states that
analyticity regions in the parameter space of ordinary differential equations (ODEs) are normally inherited by solutions of these equations.
Analytic functions are determined by their singularities, so one can talk also about singularities instead of analyticity regions.
It is not difficult to see that the unconstrained HE equations are singular at
$G = 0$ or, equivalently, at $ \WI{\lambda}{G} = \infty $.
According to the Poincar\'e theorem, the general solution
is thus expected to be singular at $ \WI{\lambda}{G} = \infty $ as well.
One of the unconstrained HE equations becomes an LEC in the limit of $ \WI{\lambda}{G} = \infty $.
It would be natural to consider the LEC solution as a zeroth-order approximation
of the general solution in the framework of perturbation theory in the small parameter $1/\WI{\lambda}{G}$.
The Poincar\'e theorem suggests, however, that as long as $ \WI{\lambda}{G} = \infty $ is a singular point, such an expansion fails
because the radius of convergence of the perturbation series is zero.
Dyson's argument is interpreted similarly, with the same conclusions for convergence of the perturbation series.

The general solution to the unconstrained HE equations is therefore expected to be nonperturbative,
which implies that the density distributions of the fluid components are singular at $ \WI{\lambda}{G} = \infty $.
Such solutions are typical of singularly perturbed ODE systems (see, e.g., \cite{Malley1990})
whose order changes in the limit of a vanishingly small parameter.
The unconstrained HE equations belong to this class of singularly perturbed ODE systems,
with the small parameter being $1/ \WI{\lambda}{G}$.

In this paper, the properties of the general solution to the unconstrained HE equations are investigated within the NGT framework.
The outline of the paper is as follows:  Section II presents the unconstrained HE equations
for a two-component fluid of ions and electrons described by polytropic equations of state (EoS).
A unique regular solution at $ \WI{\lambda}{G} = \infty $,
which is a particular solution to these equations, is analyzed in detail.
A regular solution can be constructed using perturbation theory in the small parameter $ 1 / \WI{\lambda}{G} $, starting from the LEC solution.
The solutions to the unconstrained HE equations that have previously been discussed in Refs.~\cite{Rosseland1924} -- \cite{Beznogov2013}
are identified as such regular solutions.
The regular solutions are investigated
(i) for equal polytropic indices of ions and electrons,
(ii) by expanding the solution in a power series in the stellar radius near the center, and
(iii) by considering an expansion in $1/\WI{\lambda}{G}$ around the LEC solution.
In Sect.~III, the electron shell of a star is discussed.
In Sect.~IV, we consider an exactly solvable model with unit polytropic indices of ions and electrons,
which perfectly illustrates the non-perturbative nature of the general solution
and clarifies the important role of the irregular part of the general solution in the formation of the electro- or ionospheres of stars.
The mass, radius and charge of a star are found as functions of the central baryon number and charge densities.
The exponential sensitivity of the electro- and ionospheric characteristics to the central charge density is established.
A useful example of an exactly solvable model of a
charged star consisting of a two-component incompressible fluid
is further discussed in Appendix A.
In Sect.~V, we construct the general solution to the unconstrained HE equations for arbitrary polytropic indices.
In the central stellar region, the irregular, non-perturbative part of the general solution can be found using
perturbation theory of singular ODE systems (see, e.g., \cite{Malley1990}).
Such problems arise in the semiclassical limit of the Schr\"odinger equation,
where the method of Wentzel, Kramers and Brillouin, commonly known as the WKB approximation,
is proven to be effective.
Perturbation theory applies everywhere, except in a thin subsurface layer of the star with a thickness of $\sim 1/\sqrt{\WI{\lambda}{G}}$.
To match the internal solution to the outer electron or ion shell, numerical methods can be used.
The conclusions section summarizes the results.

%%%%%%%%%%%%%%%%%%%%%%%%%%%%%%%%%%%%%%%%%%%%%%%%%%%%%%%%%%%%%%%%%
\section{Regular solution to the unconstrained hydrostatic equilibrium equations}
%%%%%%%%%%%%%%%%%%%%%%%%%%%%%%%%%%%%%%%%%%%%%%%%%%%%%%%%%%%%%%%%%
\setcounter{equation}{0}
\noindent

We consider the problem of the polarization of a substance in a spherically symmetric star.
The thermodynamic equilibrium equations for the chemical potentials of the ions, $ \WI{\mu}{i} $, and the electrons, $ \WI{\mu}{e} $,
coincide with Gibbs' condition (\ref{Gibbs}) and can be written as
\begin{eqnarray}
\WI{\mu}{i} + \WI{m}{i}\WI{\varphi}{G} + Ze \WI{\varphi}{E} &=& \mathrm{const}, \label{mu_i_eq} \\
\WI{\mu}{e} + \WI{m}{e}\WI{\varphi}{G} - e \WI{\varphi}{E}  &=& \mathrm{const}, \label{mu_e_eq}
\end{eqnarray}
where $\WI{\varphi}{G}$ and $\WI{\varphi}{E}$ are the gravitational and electrostatic potentials,
$ eZ $ and $\WI{m}{i} = A m_\mathrm{u} $ are the charge and mass of the ions.
Bearing in mind possible applications to physics of white dwarfs,
we set temperature equal to zero, and suppose the dominance of the contribution of electrons in thermodynamic functions.
The chemical equilibrium with respect to the weak interactions is neglected.
We also assume the ion and electron EoS are polytropic with $ P \sim  n^{\gamma}$
and attempt to reduce the problem to one known in the theory of polytropes (see, e.g., Ref.~\cite{Horedt2004}).

%%%%%%%%%%%%%%%%%%%%%%%%%%%%%%%%%%%%%%%%%%%%%%%%%%%%%%%%%%%%%%%%%
\subsection{Two-fluid model}
%%%%%%%%%%%%%%%%%%%%%%%%%%%%%%%%%%%%%%%%%%%%%%%%%%%%%%%%%%%%%%%%%

The gravitational and electrostatic potentials %${\varphi}_{G}$ and ${\varphi}_{E}$
satisfy the Poisson equations
\begin{eqnarray}
\triangle\WI{\varphi}{G} &=&  4\pi G\WI{\rho}{m},\label{Poisson_G}\\
\triangle\WI{\varphi}{E} &=& -4\pi \WI{\rho}{e},\label{Poisson_E}
\end{eqnarray}
where the mass and charge densities are, respectively,
\begin{eqnarray}
\WI{\rho}{m}&=&  \WI{m}{i} \WI{n}{i} + \WI{m}{e} \WI{n}{e},\label{rho_m} \\
\WI{\rho}{e}&=& Ze\WI{n}{i}-e \WI{n}{e}. \label{rho_e}
\end{eqnarray}
The radial derivative term of the Laplacian $\triangle$ is equal to
\begin{equation*}
\triangle_r = \frac{1}{r^2}\frac{d}{dr}\left(r^2\frac{d}{dr}\right).\label{Laplasian}
\end{equation*}

By applying $\triangle$ to Eqs.~(\ref{mu_i_eq}) and (\ref{mu_e_eq}) and constructing appropriate linear combinations of the chemical potentials,
we obtain
\begin{eqnarray}
\triangle_{r}(Z \WI{\mu}{e}+ \WI{\mu}{i})&=&-4\pi G (\WI{m}{i}{+}Z\WI{m}{e})\WI{\rho}{m},                   \label{main1}\\
\triangle_{r}( \WI{m}{i} \WI{\mu}{e}-\WI{m}{e} \WI{\mu}{i})&=&-4\pi e (\WI{m}{i}{+}Z\WI{m}{e})\WI{\rho}{e}.\label{main2}
\end{eqnarray}

According to Eqs.~(\ref{mu_i_eq}) and (\ref{mu_e_eq}), the gradient $-\nabla (Z\WI{\mu}{e}+\WI{\mu}{i})$ %on the other hand
is equal to the gravitational force acting on an ion and $Z$ electrons.
The thermodynamic relationship $ dP = n d \mu $ allows the gradients of the chemical potentials to be presented in terms of pressure gradients.
As a special case, the hydrostatic limit of the Euler equation is recovered for an electroneutral fluid with $ \WI{n}{e} = Z \WI{n}{i} $:
\begin{equation*}
\frac{1}{\WI{\rho}{m}}\nabla(\WI{P}{i}+\WI{P}{e}) + \nabla \WI{\varphi}{G} = 0.\label{Pgrad}
\end{equation*}
As mentioned above, we assume that $ \WI{P}{i} \ll \WI{P}{e} $ holds everywhere.

First, we define the standard relations between pressures and concentrations from the theory of polytropes:
\begin{equation}
\WI{P}{k} = \WI{K}{k} \WI{n}{k}^{1 + 1/{\etak{k}}}, \quad \mathrm{k} = \mathrm{(i,e)},
\end{equation}
where $\WI{K}{k}$ and $\etak{k}$ are fixed constants and $\etak{k}$ is called the polytropic index.
The ion and electron components interact with each other only through the gravitational and electrostatic forces.
No fundamental difficulties arise when considering the influence of the interactions of the components (see, e.g., \cite{Yoselevsky2009}), although this consideration requires special attention.
The dimensionless functions $ \WI{\theta}{k} $ are defined as follows
\begin{equation}
\WI{n}{k}\equiv \WI{n}{k0} \WI{\theta}{k}^{\etak{k}}, \label{n_theta}
\end{equation}
where the $\WI{n}{k0}$ are the concentrations at $ r = 0 $.
Excluding the rest mass, the chemical potentials are equal to $\WI{\mu}{k} = \WI{\mu}{k0} \thetak{k} $, where $\WI{\mu}{k0} = \WI{K}{k}(1 + \etak{k})\WI{n}{k0}^{1/\etak{k}}$.
The functions $ \thetak{k} =  \thetak{k} (x)$ depend on a dimensionless coordinate $ x $, which is related to $ r $ as follows:
\begin{equation}
r=r_0 x,\quad r_0^2 = \frac{Z \WI{\mu}{e0}}{4\pi G \WI{m}{i}(\WI{m}{i} + Z\WI{m}{e})\WI{n}{i0}}.\label{r0}
\end{equation}
These expressions allow Eqs.~(\ref{main1}) and (\ref{main2}) to be rewritten in the following dimensionless form:
\begin{eqnarray}
\triangle_{x}(\thetak{e}+\Lamb{i}\thetak{i}) &=& - (\thetak{i}^{\etak{i}}+\Lamb{m}\Lamb{e}\thetak{e}^{\etak{e}}),\label{main_tet_1}\\
\triangle_{x}(\thetak{e}-\Lamb{m}\Lamb{i}\thetak{i}) &=& -\Lamb{G} (\thetak{i}^{\etak{i}}-\Lamb{e}\thetak{e}^{\etak{e}}),\label{main_tet_2}
\end{eqnarray}
with the initial conditions
\begin{equation}
\thetak{k}(0)=1, ~~~~ \theta^{\prime}_{\mathrm{k}}(0)=0. \label{initial_conditions}
\end{equation}
The Laplacian $\triangle_{x}$ that appears in Eqs.~(\ref{main_tet_1}) and (\ref{main_tet_2}) acts on the coordinate $x$.
The dimensionless $\Lambda$ parameters are defined below.

The ratio between the densities $ \WI{n}{e0}$ and $\WI{n}{i0} $ determines the parameter
\begin{equation}
\Lamb{e}=\frac{\WI{n}{e0}}{Z\WI{n}{i0}} \approx 1,\label{Lambdae}
\end{equation}
which characterizes the deviation from electroneutrality in the center of the star.
In the limit of $\Lamb{G} = \infty$, Eqs.~(\ref{main_tet_2}) and (\ref{initial_conditions})
require $ \Lamb{e} = 1$, which is the LEC at $r=0$.

$\Lamb{i}$ characterizes the smallness of the ion pressure:
\begin{equation}
\Lamb{i}=\frac{\WI{\mu}{i0}}
{Z \WI{\mu}{e0}}
\approx \frac{(1 + \etak{i})\WI{P}{i0}}{(1 + \etak{e}) \WI{P}{e0}} \ll 1, \label{Lambdai}
\end{equation}
where $\WI{P}{k 0} $ is the pressure of the components at the center of the star.
The last equality in (\ref{Lambdai})
is valid to the same accuracy as the approximation $ \Lamb{e} \approx 1$.
For $ \etak{e} \neq \etak{i} $,
the parameters $\Lamb{e} $ and $\Lamb{i} $ are independent.

The parameter $ \Lamb{m}$ is a measure of the smallness of the electron mass relative to the ion mass:
\begin{equation}
\Lamb{m}=\frac{Z\WI{m}{e}}{\WI{m}{i}}=\frac{Z\WI{m}{e}}{A \WI{m}{u}} \approx 5.5 \times 10^{-4} \frac{Z}{A}.\label{Lambdam}
\end{equation}

The main parameter $\Lamb{G}$ is expressed as follows:
\begin{equation}
\Lamb{G}=\frac{Z^2e^2}{G\WI{m}{i}^2} = \left(\frac{Z}{A}\right)^2 \WI{\lambda}{G} .\label{LambdaG}
\end{equation}
A high value of $\WI{\Lambda}{G}$
plays the predominant role in specifying the nature of the problem.

Notably, $ \Lamb{e} $ and $ \Lamb{i} $ depend on the EoS and the stellar structure through $\WI{n}{k0} $ and $ \WI{K}{k} $,
whereas $ \Lamb{m} $ and $ \Lamb{G} $ depend only on the chemical composition of the fluid ($ A $ and $ Z $).

Equations (\ref{main_tet_1}) -- (\ref{initial_conditions}) constitute the system of unconstrained HE equations in dimensionless form.
This system defines a Cauchy problem with the initial conditions given in Eq.~(\ref{initial_conditions}).
In Eq.~(\ref{main_tet_2}), the small parameter $1/\Lamb{G}$ appears as a multiplier of the Laplacian; consequently,
Eqs.~(\ref{main_tet_1}) and (\ref{main_tet_2}) belong to the class of singularly perturbed ODEs.

%%%%%%%%%%%%%%%%%%%%%%%%%%%%%%%%%%%%%%%%%%%%%%%%%%%%%%%%%%%%%%%%%

%%%%%%%%%%%%%%%%%%%%%%%%%%%%%%%%%%%%%%%%%%%%%%%%%%%%%%%%%%%%%%%%%
\subsection{Two-fluid model with equal polytropic indices}
%%%%%%%%%%%%%%%%%%%%%%%%%%%%%%%%%%%%%%%%%%%%%%%%%%%%%%%%%%%%%%%%%

The key aspect of the problem is best illustrated by first considering
the case of identical polytropic indices $\etak{i} = \etak{e} \equiv \eta $.
A particular solution to the system defined in Eqs.~(\ref{main_tet_1}) and (\ref{main_tet_2})
can be represented as
$\thetak{i} (x) = \thetak{e}(x) \equiv \theta (\tilde {x}) $, with
\begin{equation}
\tilde{x}=x\sqrt{\frac{1+\Lamb{m}\Lamb{e}}{1+\Lamb{i}}},  \label{ni_eq_ne_x}
\end{equation}
where the function $\theta(\tilde{x})$ is a solution to the Lane-Emden equation
\begin{equation}
\triangle_{\tilde{x}} \theta(\tilde{x}) = -\theta^\eta (\tilde{x}) \label{Lane-Emden}
\end{equation}
with the standard initial conditions $\theta(0)=1$ and $\theta'(0)=0$.
In the case of $\etak{i} = \etak{e}$, the parameters $ \Lamb{e} $ and $ \Lamb{i} $ are dependent:
\begin{equation} \label{eta eq eta}
\Lamb{i}=\frac{\WI{K}{i}}{\WI{K}{e} \Lamb{e}^{1/\eta} Z^{1 + 1/\eta}}.
\end{equation}
The self-consistency of Eqs.~(\ref{main_tet_1}) and (\ref{main_tet_2}) requires $\Lamb{e} = \Lamb{e}^{\mathrm{reg}}$, where
\begin{equation}
\Lamb{e}^{\mathrm{reg}}=\frac{\Lamb{G}(1{+}\Lamb{i})-(1{-}\Lamb{m}\Lamb{i})}{\Lamb{G}(1{+}\Lamb{i})+\Lamb{m}(1{-}\Lamb{m}\Lamb{i})}. \label{ni_eq_ne_Lambe}
\end{equation}
Equations~(\ref{eta eq eta}) and (\ref{ni_eq_ne_Lambe}) define $ \Lamb{i} $ and $\Lamb{e} $
as functions of $\Lamb{G}$.

In the model considered here, the ion and electron densities appear to be similar, as in the Rosseland model \cite{Rosseland1924}.
The parameter $ \Lamb{e} $ differs from unity, so the LEC is violated at $ O(1/\Lamb{G}) $.
The solution describes a uniformly polarized star with a positive electric charge.

The concentrations $ \WI{n}{i0} $ and $ \WI{n}{e0} $ are considered as input parameters of the model.
One would expect the general solution to represent a two-parameter set of functions.
The constraint $\Lamb{e} = \Lamb{e}^{\mathrm{reg}}$ restricts the freedom to a one-parameter set.
This seeming paradox is explained in Sects. IV and V. Here, we merely remark that an additional
degree of freedom due to variations of $\Lamb{e}$
arises from the singular component of the general
solution to Eqs.~(\ref{main_tet_1}) and (\ref{main_tet_2}).

%%%%%%%%%%%%%%%%%%%%%%%%%%%%%%%%%%%%%%%%%%%%%%%%%%%%%%%%%%%%%%%
\subsection{A power-series expansion near the center}
%%%%%%%%%%%%%%%%%%%%%%%%%%%%%%%%%%%%%%%%%%%%%%%%%%%%%%%%%%%%%%%

A series expansion in powers of $x$ is effective for solving Eqs.~(\ref{main_tet_1}) and (\ref{main_tet_2}) for arbitrary polytropic indices,
and it sheds further light on the origin of the constraint $\Lamb{e} = \Lamb{e}^{\mathrm{reg}}$.
Considering the initial conditions, we search for a solution in the following form:
\begin{equation}
\thetak{k}=1+\sum\limits_{p=1}^{\infty}\beta_{\mathrm{k} p}x^{2p}, \quad \mathrm{k=(i,e)}.\label{theta_ie_expansion}
\end{equation}
Terms with odd powers of $ x $ are absent because of the even parity of $ \Delta_{x} $.
By substituting these expressions into Eqs.~(\ref{main_tet_1}) and (\ref{main_tet_2}) and equating the coefficients at the same powers of $ x $,
we obtain an infinite system of algebraic equations.

The zeroth-order terms yield
\begin{eqnarray}
6(\WI{\beta}{e1}+\Lamb{i} \WI{\beta}{i1})&=&-(1+\Lamb{m}\Lamb{e}),\label{m1_por0}\\
6(\WI{\beta}{e1}-\Lamb{m} \Lamb{i} \WI{\beta}{i1})&=&-\Lamb{G}(1-\Lamb{e}).\label{m2_por0}
\end{eqnarray}
The left-hand side of Eq.~(\ref{m2_por0}) is of $O(1)$,
so we require $ \Lamb{e} = 1 - \WI{\alpha}{e} / \Lamb{G} $ with $\WI{\alpha}{e} = O(1)$.
$\WI{\beta}{e1} $ and $\WI{\beta}{i1} $ can be found from Eqs.~(\ref{m1_por0}) and (\ref{m2_por0}), whereas $\WI{\alpha}{e} $ remains a free parameter.

The $\sim x^2$ terms yield the equations
\begin{eqnarray*}
20( \WI{\beta}{e2}+\Lamb{i}\WI{\beta}{i2})&=&-(\WI{\beta}{i1}\etak{i}+\Lamb{m}\Lamb{e}\WI{\beta}{e1}\etak{e}),\label{m1_por2}\\
20( \WI{\beta}{e2}-\Lamb{m}\Lamb{i}\WI{\beta}{i2})&=&-\Lamb{G}(\WI{\beta}{i1}\etak{i}-\Lamb{e}\WI{\beta}{e1}\etak{e}).\label{m2_por2}
\end{eqnarray*}
Again the boundedness of the expansion coefficients in last of the above equations yields
\begin{equation}
\WI{\beta}{i1}\etak{i1}=\WI{\beta}{e1}\etak{e1} + \frac{\gamma_1}{\Lamb{G}},\label{por2}
\end{equation}
where $\gamma_1  = O(1)$ is an unknown parameter.
This equality, with the use of Eqs.~(\ref{m1_por0}) and (\ref{m2_por0}), allows one to determine the lowest-order parameters,
including $\alpha_{e} $, to an accuracy of $ O(1/ \Lambda_{G}) $:
\begin{eqnarray*}
\frac{\WI{\beta}{e1}}{\etak{i}} &=& \frac{\WI{\beta}{i1}}{\etak{e}} = -\frac{1 + \Lamb{m}}{6(\etak{i} + \Lamb{i}\etak{e})}, \nonumber \\
%\beta_{i1}=-\frac{\eta_{e}(1 + \Lambda_m)}{6(\eta_{i} + \Lambda_i\eta_{e})},\quad
\WI{\alpha}{e}&=&(\etak{i} {-} \Lamb{m}\Lamb{i}\etak{e})\frac{1 + \Lamb{m}}{\etak{i} + \Lamb{i}\etak{e}}.\label{x_small_asympt}
\end{eqnarray*}
In the limit of $ \etak{i} = \etak{e} $, $\WI{\alpha}{e} $ is consistent with Eq.~(\ref{ni_eq_ne_Lambe}).
The higher-order terms in $ 1/ \Lamb{G} $ are determined, in turn, from the condition of the boundedness of the expansion coefficients at $ x^4 $, etc.
The regularity condition at $ \Lamb{G} = \infty$ thus imposes a constraint on the concentrations of ions and electrons in the center of the star.
We are left with a set of one-parameter solutions.

The results of this subsection can be generalized to GRT \cite{Yudin2018}.

%%%%%%%%%%%%%%%%%%%%%%%%%%%%%%%%%%%%%%%%%%%%%%%%5
\subsection{A power-series expansion in $G$}
%%%%%%%%%%%%%%%%%%%%%%%%%%%%%%%%%%%%%%%%%%%%%%%%%%%%%%%%%%%%%%%

The occurrence of a large parameter $ \Lamb{G}$ in Eqs.~(\ref{main_tet_1}) and (\ref{main_tet_2}) makes
it possible to search for solutions throughout the entire $ x $ range using a power-series expansion:
\begin{eqnarray}
\thetak{k} &=&\thetak{k0}+\thetak{k1}\Lamb{G}^{-1}  +O(\Lamb{G}^{-2}), \label{expansion_ie}\\
\Lamb{e}  &=&\Lamb{e0}+\Lamb{e1}\Lamb{G}^{-1}+O(\Lamb{G}^{-2}).
\end{eqnarray}
To a zeroth-order approximation, the regularity of the right-hand side of Eq.~(\ref{main_tet_2}) yields
$ \thetak{i0}^{\etak{i}} (x) = \Lamb{e0} \thetak{e0}^{\etak{e}} (x) $.
The initial conditions $ \thetak{e0} (0) = \thetak{i0} (0) = 1 $ imply that
\begin{equation}
\thetak{i0}^{\etak{i}}(x)\equiv\thetak{e0}^{\etak{e}}(x),\quad \Lamb{e0}=1.\label{thetas0}
\end{equation}
Equation~(\ref{main_tet_1}) yields
\begin{eqnarray}
\triangle_{x}(\thetak{e0}+\Lamb{i}\thetak{i0})&=&
-(1{+}\Lamb{m})\thetak{e0}^{\etak{e}} \nonumber \\
&=&-(1{+}\Lamb{m})\thetak{i0}^{\etak{i}}, \label{theta_main0}
\end{eqnarray}
which is the generalized polytropic equation. We solve it for the initial conditions (\ref{initial_conditions})
in the interval from $x = 0$ up to the stellar surface at $x = \WI{x}{b} $, where
$ \thetak{e0} (\WI{x}{b}) = \thetak{i0} (\WI{x}{b}) = 0 $.

The first-order terms in Eqs.~(\ref{main_tet_1}) and (\ref{main_tet_2}) yield
%\begin{widetext}
\begin{eqnarray}
&&\triangle_{x}(\thetak{e1}+\Lamb{i}\thetak{i1})= \label{first_order1} \\
&&\;\;\;\;\;\;\;\;-\thetak{e0}^{\etak{e}}\left(\etak{i}\frac{\thetak{i1}}{\thetak{i0}} + \Lamb{m}\etak{e}\frac{\thetak{e1}}{\thetak{e0}}+\Lamb{m}\Lamb{e1}\right),
\nonumber  \\
&&\triangle_{x}(\thetak{e0}-\Lamb{m}\Lamb{i}\thetak{i0})= \label{first_order2}\\
&&\;\;\;\;\;\;\;\; -\thetak{e0}^{\etak{e}}\left(\etak{i}\frac{\thetak{i1}}{\thetak{i0}} - \etak{e}\frac{\thetak{e1}}{\thetak{e0}}-\Lamb{e1}\right),\nonumber
\end{eqnarray}
%\end{widetext}
with the initial conditions $ \thetak{k1} (0) = \theta^{\prime}_{\mathrm{k1}} (0) = 0 $.
Let us consider Eq.~(\ref{first_order2}) for $ x \rightarrow 0 $. Considering the asymptotic behavior
of $ \thetak{k} $ (Eq.~(\ref{x_small_asympt}) for $ \WI{\beta}{k1} $ and $\WI{\alpha}{e} $),
we obtain the identity to zeroth order. Considering $ \Lamb{e1} = - \WI{\alpha}{e} $,
Eq.~(\ref{first_order2}) can be written as follows:
\begin{equation*}
\frac{(1+\Lamb{m})\Lamb{i}}{\etak{i}{+}\Lamb{i}\etak{e}}\triangle_{x}(\etak{e} \thetak{e0}-\etak{i}\thetak{i0})=-\thetak{e0}^{\etak{e}}\left(\etak{i}\frac{\thetak{i1}}{\thetak{i0}}-
\etak{e}\frac{\thetak{e1}}{\thetak{e0}}\right).\label{first_order2a}
\end{equation*}
For $ \etak{i} = \etak{e} $, this equation is satisfied by the similar distribution $ \thetak{i1} = \thetak{e1} $.
Equation~(\ref{first_order1}), when combined with Eq.~(\ref{first_order2}), can also be
written as follows:
\begin{eqnarray}
\triangle_{x}(\thetak{e1}{+}\Lamb{i}\thetak{i1}+\Lamb{m}[\thetak{e0}{-}\Lamb{m}\Lamb{i}\thetak{i0}])=\;\;\;\;\;\;\;\;\;\;\;\;&& \label{first_order1a} \\
-(1{+}\Lamb{m})\thetak{e0}^{\etak{e}}\etak{i}\frac{\thetak{i1}}{\thetak{i0}}.&&\nonumber
\end{eqnarray}

The zeroth- and first-order approximations are defined for all values of $ x $.
The region near the surface of the star should be treated separately,
because for $ \thetak{k0} \rightarrow 0 $, the validity of the expansion (\ref{expansion_ie}) can be violated.

%%%%%%%%%%%%%%%%%%%%%%%%%%%%%%%%%%%%%%%%%%%%%%%%%%%%%%%%%%%%%%%
\subsection{Global stellar parameters}
%%%%%%%%%%%%%%%%%%%%%%%%%%%%%%%%%%%%%%%%%%%%%%%%%%%%%%%%%%%%%%%

The stellar radius (more precisely, the boundary of its baryon component) is defined as $ \WI{R}{s} = r_0 \WI{x}{b} $,
where $ r_0 $ is given by Eq.~(\ref{r0}) and $ \WI{x}{b} $ is fixed by the condition $ \theta (\WI{x}{b}) = 0 $
from the solution to Eqs.~(\ref{main_tet_1}) and (\ref{main_tet_2}). The mass of the star takes the form
\begin{eqnarray}
\WI{M}{s}&=&\int\limits_0^{\WI{R}{s}} 4\pi r^2(\WI{n}{i}\WI{m}{i} + \WI{n}{e}\WI{m}{e}) dr \label{Mass}  \\
     &=&-4\pi r_0^3 \WI{n}{i0}\WI{m}{i} x^2 \frac{d}{dx} \left(\thetak{e}{+}\Lamb{i}\thetak{i}\right)\Big|_{x=\WI{x}{b}},\nonumber
\end{eqnarray}
where Eq.~(\ref{main_tet_1}) is used to proceed to the second line.
The total stellar charge (at the boundary of the baryon component) is equal to
\begin{eqnarray}
\WI{Q}{s}&=&\int\limits_0^{\WI{R}{s}} 4\pi r^2(Ze\WI{n}{i}{-}e\WI{n}{e})dr \label{Charge}\\
&=&-4\pi r_0^3 \WI{n}{i0}\frac{Ze}{\Lamb{G}} x^2 \frac{d}{dx}\!\left(\thetak{e}{-}\Lamb{m}\Lamb{i}\thetak{i}\right)\Big|_{x=\WI{x}{b}}, \nonumber
\end{eqnarray}
where we have used Eq.~(\ref{main_tet_2}). The stellar charge can be estimated as
\begin{equation}
\WI{Q}{s}\sim\frac{Ze\WI{M}{s}}{A m_{\mathrm{u}} \Lamb{G}}\simeq 10^{21} e \frac{A}{Z}\frac{\WI{M}{s}}{M_\odot},\label{Q_order_of_magnitude}
\end{equation}
where $ M_\odot $ is the mass of the Sun.

The total uncompensated electric charge of a star of one solar mass, on the boundary of its baryon component,
is on the order of 100 C, in agreement with the estimates of Pikel'ner \cite{Pikelner1964}, Bally and Harrison \cite{Bally1978} and Neslu\v san \cite{Neslusan2001}.
The maximum charges of neutron stars and white dwarfs are estimated to be 50 C and 500 C, respectively \cite{Olson1978}.
For comparison, one mole of $^{12}$C contains $ 6 N_{A} $ protons, whose total charge is $ 6 \times 10^{5} $ C.
The total negative charge of the Earth at its surface is estimated to be $ (4 - 5.7) \times 10^{5} $ C \cite{Chavalier2007}.
In absolute value, the solar charge is less than $ (0.4 - 1) \times 10^{18} $ C \cite{Iorio2012}.

%%%%%%%%%%%%%%%%%%%%%%%%%%%%%%%%%%%%%%%%%%%%%%%%%%%%%%%%%%%%%%%
%%%%%%%%%%%%%%%%%%%%%%%%%%%%%%%%%%%%%%%%%%%%%%%%%%%%%%%%%%%%%%%
\section{Electrosphere}
%%%%%%%%%%%%%%%%%%%%%%%%%%%%%%%%%%%%%%%%%%%%%%%%%%%%%%%%%%%%%%%
%%%%%%%%%%%%%%%%%%%%%%%%%%%%%%%%%%%%%%%%%%%%%%%%%%%%%%%%%%%%%%%
\setcounter{equation}{0}

In Sect. II, we have seen that
the self-consistent regular solution to the unconstrained HE equations %at $\Lambda_G = \infty$
describes a star with a positive charge. To move an electron from the surface of such a star
to infinity, it is necessary to perform work against the forces of gravity and Coulomb attraction.
In fact, a star can confine more electrons than it should be able to according to the regular solution.
This possibility leads to the formation of an electrosphere that is quite independent of
the internal structure of the star. We therefore discuss the electrosphere separately.

Under normal conditions, a metal is surrounded by a double charged layer, where an excess of positive charge at the subsurface
is created by the atomic lattice and a negative charge outside is created by electrons.
The thickness of the layer is equal to several Bohr radii, and the work function is several eV. An electrosphere surrounds
the surface of hypothetical strange stars \cite{Witten1984,Hu2002}. Within the strange star, the positive charge of the baryon component
is almost completely neutralized by electrons, whereas it is only partly neutralized in the subsurface.
As a result, an electron shell forms above the surface, with a thickness exceeding hundreds of fm, and
the chemical potential of electrons near the surface is several tens of MeV.
The electrons also give rise to an electrostatic potential jump across the surface separating different
phases of nuclear matter \cite{Krivoruchenko1995}.

Earlier, we presented arguments for the existence of stars with a total charge of $-0.1$ to $150$ C per solar mass,
which are able to support either an electrosphere or an ionosphere.
Here, we restrict ourselves to a discussion of the electrosphere. %Later,
Section IV describes the properties of the ionosphere.

\subsection{A polytropic model}

The regular solution explicitly yields a positive charge excess at $r=0$. %in the center of the star.
Furthermore, the balance of the gravitational and electrostatic forces ensures that the charge excess must be positive everywhere
from the central region up to the surface.
The electron component extends somewhat beyond the ion component and compensates for the stellar charge.

Applying the Laplacian to Eq.~(\ref{mu_e_eq}), we obtain
\begin{equation}
\triangle_{r} \WI{\mu}{e}=4\pi e^2\WI{n}{e}\left(1-\frac{G \WI{m}{e}^2 }{e^2}\right).\label{atm_mu}
\end{equation}
At the surface ($ r = \WI{R}{s} $), an equilibrium condition holds:
\begin{equation}
-\frac{1}{\WI{n}{e}}\frac{d \WI{P}{e}}{dr}=\frac{e\WI{Q}{s}}{\WI{R}{s}^2}+
\frac{G\WI{m}{e}\WI{M}{s}}{\WI{R}{s}^2},\label{gradP}
\end{equation}
where $ \WI{M}{s} $ and $ \WI{Q}{s} $ are given by Eqs.~(\ref{Mass}) and (\ref{Charge}).

Proceeding in the standard way, we define a dimensionless polytropic function based on the equation
$ \WI{n}{e} = \WI{n}{a0} \thetak{a}^{\etak{e}} $, where $ \WI{n}{a0} $ is the electron density at the boundary
of the ion component.
The thickness of the envelope is assumed to be much smaller than $ \WI{R}{s} $.
This assumption will be checked at the end of the calculation. The Laplacian simplifies to $ \triangle_{r} \rightarrow {d^2} / {dr^2} $.
For convenience, we introduce the dimensionless parameter
\begin{equation}
\WI{\lambda}{m} = \frac{\WI{m}{e}}{m_{\mathrm{u}}}. \label{Lamb_G_m}
\end{equation}
The dimensionless variable $ y $ is defined by
\begin{equation}
r=\WI{R}{s} + \WI{r}{a}y,\quad \WI{r}{a}^2 = \frac{\WI{\mu}{a0}}{4\pi e^2 \left(1 {-} \WI{\lambda}{m}^2/\WI{\lambda}{G}\right)\WI{n}{a0}}, \label{ra}
\end{equation}
where $\WI{\mu}{a0}$ is the chemical potential of electrons at the surface. Thus, Eq.~(\ref{atm_mu}) becomes
\begin{equation}
\frac{d^2\thetak{a}}{dy^2}=\thetak{a}^{\etak{e}}(y).\label{d2theta}
\end{equation}
At the surface, $ y = 0 $ and $ \thetak{a}(0) = 1 $.
Once the stellar structure equations are solved and the stellar mass, radius, and charge are determined,
the initial condition for $\thetak{a}^{\prime} (0)$ follows from Eq.~(\ref{gradP}).
Considering Eq.~(\ref{Q_order_of_magnitude}), we write the charge in the form
\begin{equation}
\WI{Q}{s}=\frac{\WI{M}{s}}{m_{\mathrm{u}}} \frac{e\WI{q}{s}}{\WI{\lambda}{G}},\label{dQ}
\end{equation}
where $\WI{q}{s} = O(1)$ can be found by solving the unconstrained HE equations. We remark that ${M_{\odot}}/{m_{\mathrm{u}}} \sim (\WI{\lambda}{G}/\alpha)^{3/2}$,
where $\alpha = e^2/(\hbar c) \approx 1/137$ is the fine structure constant,
and therefore, $\WI{Q}{s} \sim e \sqrt{\WI{\lambda}{G}}/\alpha^{3/2} $.
The second initial condition can be written as
\begin{equation}
\frac{d\thetak{a}}{dy}\Big|_{y=0}=-(\WI{q}{s}+\WI{\lambda}{m})
\sqrt{\frac{\WI{P}{un}}{(1 + \etak{e})\WI{P}{a0}(\WI{\lambda}{G} - \WI{\lambda}{m}^2)}}, \label{dtheta0}
\end{equation}
where $\WI{P}{a0}=\WI{K}{e} \WI{n}{a0}^{1 + 1/\etak{e}}$ is the pressure of the electrons at the boundary
and $ \WI{P}{un} $ is the natural unit of pressure in the star:
\[
\WI{P}{un} = \frac{G \WI{M}{s}^2} {4 \pi \WI{R}{s}^4}.
\]
Integrating Eq.~(\ref{d2theta}) yields
\begin{equation}
\frac{d\thetak{a}}{dy}=-\sqrt{2(W_0 - V(\thetak{a}))},\label{dtheta}
\end{equation}
where \begin{eqnarray*}
W_{0} &=& \frac{\thetak{a}^{\prime 2}(0)}{2} - \frac{1}{1+\etak{e}}, \\
V(\thetak{a}) &=& - \frac{\thetak{a}^{1 + \etak{e}}}{1+\etak{e}},
\end{eqnarray*}
and $\thetak{a}'(0)$ is given by Eq.~(\ref{dtheta0}).
A second integration yields
\begin{eqnarray}
y = \WI{y}{a} &-& \frac{\thetak{a}}{\sqrt{2(W_{0} - V(\thetak{a}))}} \label{theta} \\
&\times&{_2}F_1 \left(\frac{1}{2},1;\frac{2+\etak{e}}{1+\etak{e}}; \frac{- V(\thetak{a})}{W_0 - V(\thetak{a})}\right), \nonumber
\end{eqnarray}
where $_2 F_1 $ is the Gauss hypergeometric function. Equation (\ref{theta})
implicitly specifies $ \thetak{a} = \thetak{a} (y) $.
With the initial condition $ \thetak{a} (0) = 1 $,
the boundary of the electrosphere can be found from $ \thetak{a} (\WI{y}{a}) = 0 $ to give
\begin{equation*}\label{x_a}
\WI{y}{a}= \frac{1}{|\thetak{a}^{\prime}(0)|} \ _2 F_1\!\!\left(\frac{1}{2},1;\frac{2+\etak{e}}{1+\etak{e}};
\frac{2}{(1+\etak{e})\thetak{a}^{\prime 2}(0)}\right).
\end{equation*}

Figure 1 shows the behavior of $\WI{y}{a}$ as a function of $ W_0 $. With a decrease in $ W_0 $, the radial thickness of the envelope increases,
and for $ W_0 \to 0 $, the radial thickness goes to infinity. However, the function $ \WI{y}{a} $ is still well defined at zero:
\begin{equation}
\WI{y}{a}(W_0 = 0) = - \frac{\sqrt{2(\etak{e}{+}1)}}{\etak{e}{-}1} < 0,
\label{yaW0}
\end{equation}
because of the irregularity of %(\ref{x_a})
$\WI{y}{a}$ for $ W_0 = 0 $. In comparison with the nonrelativistic case,
the shell of relativistic electrons extends higher.

%%%%%%%%%%%%%%%%%%%%%%%%%%%%%%%%%%%%%%%%%%%%%%%%%%%%%%%%%%%%%%%%%%%%%%%%%%%%
%%%%%%%%%%%%%%%%%%%%%%%%%%%%%%%%%%%%%%%%%%%%%%%%%%%%%%%%%%%%%%%%%%%%%%%%%%%%
%%%%%%%%%%%%%%%%%%%%%%%%%%%%%%%%%%%%%%%%%%%%%%%%%%%%%%%%%%%%%%%%%%%%%%%%%%%%
\begin{figure} [t] %
\begin{center}
\hspace{ 4mm}
\includegraphics[angle = 0,width=0.45\textwidth]{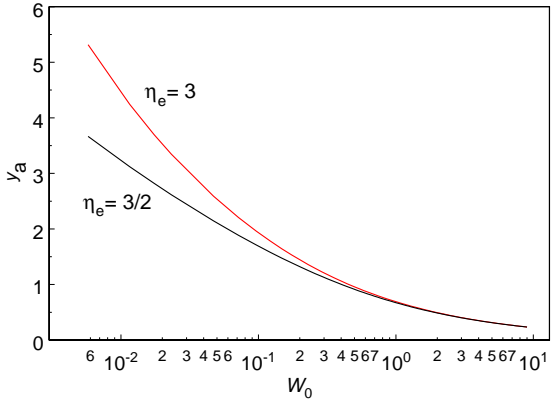}
\hspace{-4mm}
\vspace{2mm}
\caption{(color online)
Electron shell thickness $\WI{y}{a}$ in the nonrelativistic ($ \etak{e} = 3/2 $) and relativistic ($ \etak{e} = 3$) cases versus the parameter $ W_0 $.}
\label{atmos}
\end{center}
\end{figure}
%%%%%%%%%%%%%%%%%%%%%%%%%%%%%%%%%%%%%%%%%%%%%%%%%%%%%%%%%%%%%%%%%%%%%%%%%%%%
%%%%%%%%%%%%%%%%%%%%%%%%%%%%%%%%%%%%%%%%%%%%%%%%%%%%%%%%%%%%%%%%%%%%%%%%%%%%
%%%%%%%%%%%%%%%%%%%%%%%%%%%%%%%%%%%%%%%%%%%%%%%%%%%%%%%%%%%%%%%%%%%%%%%%%%%%

Equation (\ref{d2theta}) can be interpreted as the equation of motion of a point particle of unit mass in a potential $ V(\thetak{a}) $,
where $ y $ plays the role of a time variable. Equation (\ref{dtheta}) follows from the conservation of the total energy:
\[
W = \frac{\thetak{a}^{\prime 2}}{2} + V(\thetak{a}) .
\]
There are three types of solutions. If the initial momentum is small, the particle moves to the left relative to its initial position $ \thetak{a} (0) = 1 $,
stops at some point $0 < \thetak{a} < 1 $, and then accelerates in the positive direction and goes to infinity.
The total energy of the particle is negative: $ W = W_0 <0 $. In the Thomas-Fermi theory of multielectron atoms, solutions of this type are called \textit{compressed};
they terminate at $ \thetak{a}^{\prime} = 0 $ for a finite electron density,
for which the global electroneutrality of the atom holds \cite{LandauV3}.
In our problem, however, such solutions are not physical since the stars are, of course, isolated.
The second type of solution corresponds to the case of zero energy: the particle arrives at the point $ \thetak{a} = 0 $ and stops there, such that $ \thetak{a}^{\prime} = 0 $.
In the Thomas-Fermi theory, such solutions describe neutral atoms. In our case, they describe stars with a maximum negative
total charge, i.e., $ Q_{\mathrm {tot}}^{\max} < 0 $.
The third type of solution is obtained for $ W_0> 0 $. The particle arrives at the point $ \thetak{a} = 0 $ with a finite velocity of $ \thetak{a}^{\prime} <0 $,
and then the solution terminates. In the Thomas-Fermi theory, solutions of this type correspond to ionized atoms with a total positive charge.
In our case, such a solution describes an electrosphere of a finite thickness $y_a$, and the total charge of the star can be negative, zero, or positive.

The existence of solutions, therefore, depends on the derivative of the chemical potential of the electrons at the boundary of the ion component.
The absolute value of the derivative is constrained from below by the condition $ W_0 \geq 0 $, which imposes an upper limit on the number
of electrons in the shell.

At the boundary of the ion component, the total positive charge of the star is equal to $ \WI{Q}{s} $, and the charge of the electrosphere $\WI{Q}{e}$ either partially, completely,
or excessively compensates for $ \WI{Q}{s} $. Considering Eqs.~(\ref{d2theta}) and (\ref{dtheta}),
the charge of the electrosphere can be written in the form
\begin{eqnarray}
\WI{Q}{e} &=& - 4\pi \WI{R}{s}^2\int_0^{\WI{y}{a}} e \WI{n}{e}\WI{r}{a} dy \label{compensate}  \\
&=& 4\pi \WI{R}{s}^2 e \WI{n}{a0}\WI{r}{a}
\left( \thetak{a}^{\prime}(0) + \sqrt{\thetak{a}^{\prime 2}(0) - \frac{2}{1 + \etak{e}}}\right). \nonumber
\end{eqnarray}
Let $ \kappa$
be a parameter characterizing the screening, such that $\WI{Q}{e} = - \kappa \WI{Q}{s}$, where $ 0 \leq \kappa \leq \kappa_{\max} $.
In dimensionless units (see Eq.~(\ref{dQ})), the charge of the electrosphere is equal to
$ \WI{q}{e} = - \kappa \WI{q}{s} $. In particular, for $ \kappa = 1 $, the screening is complete, and the total stellar charge vanishes.
From Eq.~(\ref{compensate}), one can find
\begin{equation*}
\WI{P}{a0}=\kappa \WI{q}{s} \frac{\WI{P}{un}}{\WI{\lambda}{G}}\left[\WI{q}{s} + \WI{\lambda}{m}-
\frac{\kappa \WI{q}{s}}{2}(1{-}\WI{\lambda}{m}^2/\WI{\lambda}{G})\right],\label{P_e0_beta}
\end{equation*}
implying that the pressure in the envelope is $ \WI{\lambda}{G}$ times smaller than $ \WI{P}{un} $.
It is now convenient to rewrite Eq.~(\ref{dtheta0}) in the form
\begin{widetext}
\begin{equation}
\frac{d\thetak{a}}{dy}\Big|_{y=0}=-
\frac{\WI{q}{s}+\WI{\lambda}{m}}{\sqrt{\kappa \WI{q}{s} (1 + \etak{e})(1 {-} \WI{\lambda}{m}^2/\WI{\lambda}{G})(q_{s} + \WI{\lambda}{m} -
\kappa \WI{q}{s}/2(1 {-} \WI{\lambda}{m}^2/\WI{\lambda}{G}))}}
.\label{dtheta0_dim}
\end{equation}
\end{widetext}

The maximum negative charge of the envelope corresponds to $ W_0 = 0 $. From Eq.~(\ref{dtheta0_dim}),
one can find the critical value of the compensation parameter:
\begin{equation}
\kappa_{\max} = \frac{1 + \WI{\lambda}{m}/\WI{q}{s}}{1 - \WI{\lambda}{m}^2/\WI{\lambda}{G}} > 1.
\end{equation}
Solutions with a total negative charge, $ Q_{\mathrm{tot}} =  \WI{Q}{s}+ \WI{Q}{e}= \WI{Q}{s} (1 - \kappa) <0 $, are admissible
since an electrically neutral star has a gravitational field capable of binding electrons under the condition
that the energy of their Coulomb repulsion does not exceed the energy of their gravitational attraction to the star.
An estimate of $\WI{Q}{s} \sim 100 $ C was obtained above. The maximum negative charge $ Q_{\mathrm{tot}}^{\max} $
is proportional to the parameter $ \WI{\lambda}{m} \ll 1 $; thus, $ Q_{\mathrm{tot}}^{\max} \sim - 0.05 $ C per solar mass.

The gravitational collapse of a charged single-component spherically symmetric fluid is discussed
in Refs.~\cite{Bekenstein1971,Lasky2007}.

The maximum charge of a neutron star is determined in Ref.~\cite{Ghezzi2005} on the basis of the requirement of gravitational stability.
We showed that the maximum charge can instead be determined from the requirement for the existence of a solution of finite extent
for the electro- or ionosphere of the star.
The latter requirement yields a limit that is $\sqrt{\WI{\lambda}{G}}$ times stronger than that obtained from the requirement of gravitational stability.

The positive charge excess of strange stars at the quark surface is found in Refs.~\cite{Negreiros2009,Arbanil2015} to be about $10^{20}$ C.
The electrosphere compensates this charge, although the detailed mechanism has not yet been
studied. The particular values of the bag constant $B$ used in Refs.~\cite{Negreiros2009,Arbanil2015}
rule out the existence of massive neutron stars metastable against conversion to strange stars \cite{Krivoruchenko1987,Krivoruchenko1991}.

%%%%%%%%%%%%%%%%%%%%%%%%%%%%%%%%%%%%%%%%%%%%%%%%%%%%%%%%%%%%%%%%%%%
\subsection{Electrically neutral stars}
%%%%%%%%%%%%%%%%%%%%%%%%%%%%%%%%%%%%%%%%%%%%%%%%%%%%%%%%%%%%%%%%%%%

For $W_0 = 0$, Eqs.~(\ref{theta}) - (\ref{yaW0}) yield
\begin{equation}
\thetak{a}^{\max}(y)=\left(1 + \frac{y}{y_{0}}\right)^{-\frac{2}{\etak{e}{-}1}}, \label{theta_singular}
\end{equation}
where $y_{0} = - \WI{y}{a}(W_0 = 0) > 0$.
This solution extends to infinity, which means that an electron can be moved from the center of the star to infinity without doing work.
In the case of a relativistic electron gas, $ \etak{e} = 3 $, the solution decays rather slowly: $ \thetak{a}^{\max} \sim 1/y $.
At low densities, the electrons become nonrelativistic, with $ \etak{e} = {3}/{2} $, and the solution
approaches $ \thetak{a}^{\max} \sim 1/y^{4} $.
It is not difficult to directly verify that $ \thetak{a}^{\max} (y) $ is a solution to Eq.~(\ref{d2theta}) with the proper boundary conditions.

%%%%%%%%%%%%%%%%%%%%%%%%%%%%%%%%%%%%%%%%%%%%%%%%%%%%%%%%%%%%%%%%%%%%%%%%%%%%
%%%%%%%%%%%%%%%%%%%%%%%%%%%%%%%%%%%%%%%%%%%%%%%%%%%%%%%%%%%%%%%%%%%%%%%%%%%%
%%%%%%%%%%%%%%%%%%%%%%%%%%%%%%%%%%%%%%%%%%%%%%%%%%%%%%%%%%%%%%%%%%%%%%%%%%%%
\begin{figure} [t] %
\begin{center}
\includegraphics[angle = 0,width=0.45\textwidth]{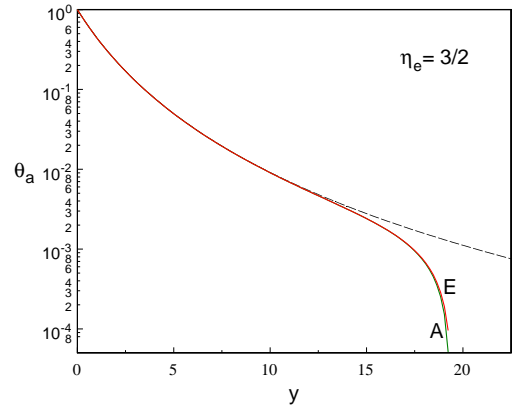}
\vspace{2mm}
\caption{(color online)
Chemical potential $\thetak{a}$ of nonrelativistic electrons
as a function of the distance from the surface of a neutral star.
The dashed curve shows the limiting solution $\thetak{a}^{\max}$ to Eq.~(\ref{theta_singular}),
the solid curve labeled E shows the exact solution to Eq.~(\ref{theta}), and the solid curve labeled A shows the approximate solution $\thetak{a}^{\max} + \sigma $.
}
\label{Pix-solution}
\end{center}
\end{figure}
%\vspace{-5cm}
%%%%%%%%%%%%%%%%%%%%%%%%%%%%%%%%%%%%%%%%%%%%%%%%%%%%%%%%%%%%%%%%%%%%%%%%%%%%
%%%%%%%%%%%%%%%%%%%%%%%%%%%%%%%%%%%%%%%%%%%%%%%%%%%%%%%%%%%%%%%%%%%%%%%%%%%%
%%%%%%%%%%%%%%%%%%%%%%%%%%%%%%%%%%%%%%%%%%%%%%%%%%%%%%%%%%%%%%%%%%%%%%%%%%%%

In view of the approximate equality $ \kappa_{\max} \approx 1 $, the electrosphere of a neutral star with $ \kappa = 1 $ is described by a solution close to
$ \thetak{a}^{\max}  (y) $.
We return to Eq.~(\ref{dtheta0_dim}) and set $ \kappa = 1 $. From this equation,
by neglecting the small terms $ \sim 1/\WI{\lambda}{G}$ and $\sim \WI{\lambda}{m}$,
the derivative of $ \thetak{a}  (y) $ at $y=0$ can be found to be
\begin{equation}
\frac{d\theta}{dy}\Big|_{y=0}\approx-\sqrt{\frac{2}{1+\etak{e}}}(1 + \xi),
\label{dtheta0_dim_ser}
\end{equation}
where
\begin{equation}
\xi=\frac{1}{2}\left(\frac{\WI{\lambda}{m}}{\WI{q}{s}}\right)^2 \ll 1.
\label{aplha}
\end{equation}
We are looking for a solution to Eq.~(\ref{d2theta}) in the form $ \thetak{a} = \thetak{a}^{\max} + \sigma $, where $\sigma $ is a small correction.
The initial conditions are of the form $ \sigma (0) = 0 $ and $ \sigma^{\prime}(0) = - \xi \sqrt{{2}/(1+\etak{e})}$.
To the lowest order, $ \sigma $ satisfies the equation
\begin{equation}
\frac{d^2\sigma}{dy^2}=\frac{\etak{e}\sigma}{\left(1+\frac{y}{y_0}\right)^2}.
\label{sigma}
\end{equation}
The solution to this equation is given by
\begin{equation*}
\sigma(y)=\frac{2\xi}{3\etak{e}+1}\left[\left(1+\frac{y}{y_0}\right)^{-\frac{\etak{e}+1}{\etak{e}-1}}-\left(1+\frac{y}{y_0}\right)^{\frac{2\etak{e}}{\etak{e}-1}}\right].
\label{sigma_sol}
\end{equation*}

%%%%%%%%%%%%%%%%%%%%%%%%%%%%%%%%%%%%%%%%%%%%%%%%%%%%%%%%%%%%%%%%%%%%%%%%%%%%
%%%%%%%%%%%%%%%%%%%%%%%%%%%%%%%%%%%%%%%%%%%%%%%%%%%%%%%%%%%%%%%%%%%%%%%%%%%%
%%%%%%%%%%%%%%%%%%%%%%%%%%%%%%%%%%%%%%%%%%%%%%%%%%%%%%%%%%%%%%%%%%%%%%%%%%%%
\begin{figure} [t] %
\begin{center}
\includegraphics[angle = 0,width=0.45\textwidth]{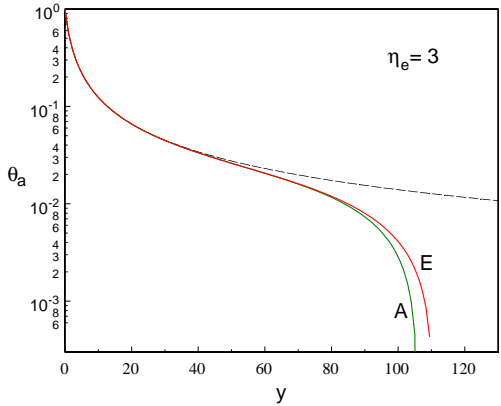}
\vspace{2mm}
\caption{(color online)
Chemical potential $\thetak{a}$ of relativistic electrons
as a function of the distance from the surface.
All parameters and notation are the same as in Fig.~\ref{Pix-solution}.
}
\label{Pix-solution-R}
\end{center}
\end{figure}
%\vspace{-5cm}
%%%%%%%%%%%%%%%%%%%%%%%%%%%%%%%%%%%%%%%%%%%%%%%%%%%%%%%%%%%%%%%%%%%%%%%%%%%%
%%%%%%%%%%%%%%%%%%%%%%%%%%%%%%%%%%%%%%%%%%%%%%%%%%%%%%%%%%%%%%%%%%%%%%%%%%%%
%%%%%%%%%%%%%%%%%%%%%%%%%%%%%%%%%%%%%%%%%%%%%%%%%%%%%%%%%%%%%%%%%%%%%%%%%%%%

The dimensionless chemical potential of the electrosphere of a neutral star
as a function of the distance to the surface is shown
in Fig.~\ref{Pix-solution} for $ \etak{e} = 3/2 $ and
in Fig.~\ref{Pix-solution-R} for $\etak{e} = 3$.
The equilibrium conditions (\ref{mu_i_eq}) and (\ref{mu_e_eq})
are valid for nonrelativistic particles. For completeness, we also consider the relativistic case $ \eta_e = 3 $,
which can be interesting for qualitative estimates. % when relativistic effects are only moderately important.
The total stellar charge is assumed to be zero ($\kappa = 1$),
whereas the parameter $ \xi = 1.5\times 10^{-7}$ corresponds to $\WI{q}{s} = 1$.
The dashed curves show the asymptotic solutions corresponding to Eq.~(\ref{theta_singular}),
the solid curves labeled E show the exact solutions corresponding to Eq.~(\ref{theta}), and the solid curves labeled A show the approximations $\thetak{a}^{\max} + \sigma $.
The approximations reproduce the exact formula in Eq.~(\ref{theta}) with reasonable accuracy.
The thickness of the electrosphere can estimated to be
\begin{equation}
\frac{\WI{y}{a}}{y_{0}} \approx  \left(\frac{3\etak{e}+1}{2\xi}\right)^{\frac{\etak{e}-1}{2(\etak{e}+1)}}-1.
\label{x_b_approx}
\end{equation}

An order-of-magnitude estimate of the thickness of the electrosphere $ \WI{y}{a} \WI{r}{a} $ can be found from Eq.~(\ref{ra}).

For nonrelativistic electrons with $ \etak{e} = 3/2 $, the pressure and density are related by $ \WI{P}{e} \sim {\hbar^2} \WI{n}{e}^{5/3} / {\WI{m}{e}} $.
Considering that $ \WI{P}{a0} \sim \WI{P}{un} / \WI{\lambda}{G} $, we obtain
\begin{eqnarray}
\frac{\WI{r}{a}}{\WI{R}{s}} &\sim&
\left[\left(\frac{\WI{a}{B}}{\WI{R}{s}}\right)^3 \frac{m_{\mathrm{u}}}{\WI{M}{s}}\WI{\lambda}{G}\right]^{1/5} \label{r0_non_rel} \\
&\approx& 2.2\times 10^{-16} \left(\frac{R_{\odot}}{\WI{R}{s}}\right)^{3/5} \left(\frac{M_{\odot}}{\WI{M}{s}} \right)^{1/5}, \nonumber
\end{eqnarray}
where $ \WI{a}{B} = \hbar^2 / (e^2 \WI{m}{e}) $ is the Bohr radius.
In comparison with the stellar radius, the thickness of the electrosphere is quite small,
although it is not small on an atomic scale:
$ \WI{y}{a} \WI{r}{a} \sim 6 \times 10^{4}\WI{a}{B}$ for an electrically neutral star of mass $ \WI{M}{s} = M_{\odot} $ and radius $ \WI{R}{s} = R_{\odot} $.

In the case of $ \etak{e} = 3 $, %, which may be of interest when relativistic effects are only moderately important.
the substitution of the relativistic relationship
$\WI{P}{e}\sim \hbar c \WI{n}{e}^{4/3}$ into Eq.~(\ref{ra}) yields
\begin{equation}
\frac{\WI{r}{a}}{\WI{R}{s}}\sim \sqrt{\frac{\WI{\lambda}{G}}{\alpha^{3/2}}\frac{m_{\mathrm{u}}}{\WI{M}{s}}}
\approx 1.3\times10^{-9}\left(\frac{M_\odot}{\WI{M}{s}}\right)^{1/2}. \label{r0_rel}
\end{equation}
The thickness of the electrosphere of an electrically neutral neutron star with a radius of $ \WI{R}{s} = 10 $ km and a mass of $ 1.4 ~ M_{\odot} $
can be estimated to be $\WI{y}{a} \WI{r}{a} \sim 1 $ mm.

%%%%%%%%%%%%%%%%%%%%%%%%%%%%%%%%%%%%%%%%%%%%%%%%%%%%%%%%%%%%%%%%%%%
%%%%%%%%%%%%%%%%%%%%%%%%%%%%%%%%%%%%%%%%%%%%%%%%%%%%%%%%%%%%%%%%%%%
\section{Exactly solvable model}
%%%%%%%%%%%%%%%%%%%%%%%%%%%%%%%%%%%%%%%%%%%%%%%%%%%%%%%%%%%%%%%%%%%
%%%%%%%%%%%%%%%%%%%%%%%%%%%%%%%%%%%%%%%%%%%%%%%%%%%%%%%%%%%%%%%%%%%
\setcounter{equation}{0}

As seen from the previous discussion,
the regular solution describes a polarized star with a nonzero positive charge as measured at the ion surface.
The properties of the electrosphere are uniquely determined by the internal structure of the star.
The electrosphere ensures that the total stellar charge is generally not equal to zero.
This feature is particularly evident in the model with equal polytropic indices considered in Sect.~II.B.
An electrosphere occurs provided that $ \WI{\theta}{e} (\WI{x}{b}) \neq 0 $, whereas for identical polytropic indices,
the distributions vanish at the same point, $ x = \WI{x}{b} $.

Earlier, we presented simple physical arguments in favor of the existence of stars with electric charges in the range of $- 0.1$ to $150$ C per solar mass.
A regular solution corresponds to a single representative of this class.
The regular solutions described in Refs.~\cite{Rosseland1924} -- \cite{Belvedere2012} and in Sect. II above
do not exhaust all physical possibilities, nor are they general solutions to the unconstrained HE equations.

Fourth-order systems of ODEs, similar to Eqs.~(\ref{main_tet_1}) and (\ref{main_tet_2}), depend on four initial conditions.
In our case, two of the initial conditions constrain the derivatives of the chemical potentials (cf. Eq.~(\ref{initial_conditions})).
These conditions ensure smooth behavior of the solutions at $r=0$ and cannot be modified.
The remaining two initial conditions fix $\WI{\mu}{i}$ and $\WI{\mu}{e}$ and thereby define a two-parameter set of stellar configurations.
As shown in Sect.~II, the requirement of regularity imposes a constraint on the value of $\WI{\Lambda}{e}$,
such that only a one-parameter set of stellar configurations survives.

Some of the solutions to Eqs.~(\ref{main_tet_1}) and (\ref{main_tet_2}) appear to be lost
with the imposition of a requirement for regularity in the neighborhood of
$ \WI{\Lambda}{G} = \infty $.

%%%%%%%%%%%%%%%%%%%%%%%%%%%%%%%%%%%%%%%%%%%%%%%%%%%%%%%%%%%%%%%%%%%
\subsection{General properties of solutions}
%%%%%%%%%%%%%%%%%%%%%%%%%%%%%%%%%%%%%%%%%%%%%%%%%%%%%%%%%%%%%%%%%%%

The analytic properties of solutions to ODE systems are determined by the Poincar\'e theorem on analyticity (see, e.g., \cite{Ilyashenko2008}),
which asserts, roughly speaking, that solutions to ODE systems, when they exist,
are analytic functions of the initial coordinates and parameters in the region of analyticity of the ODEs.
The origin of non-regular solutions can also be understood on the basis of Dyson's argument,
which provides an effective qualitative criterion for the non-analyticity of observables in terms of the system parameters \cite{Dyson1952}.
From these perspectives, we discuss the analytic properties in the parameter $ \WI{\Lambda}{G}$
of the solutions to Eqs.~(\ref{main_tet_1}) and (\ref{main_tet_2}).

%%%%%%%%%%%%%%%%%%%%%%%%%%%%%%%%%%%%%%%%%%%%%%%%%%%%%%%%%%%%%%%%%%%
\subsubsection{Poincar\'e theorem on analyticity}
%%%%%%%%%%%%%%%%%%%%%%%%%%%%%%%%%%%%%%%%%%%%%%%%%%%%%%%%%%%%%%%%%%%

Using the independent variables $ \WI{\pi}{k} = \thetak{k}^{\prime} $,
Eqs.~(\ref{main_tet_1}) and (\ref{main_tet_2}) can be reduced to a system of four first-order ODEs.
In terms of $ \Phi = (\WI{\pi}{e}, \thetak{e}, \WI{\pi}{i}, \thetak{i}) $,
these equations
take the form $  \Phi^{\prime} = \mathfrak {F} (x,  \Phi, \boldsymbol\Lambda) $,
where $ \mathfrak{F} $ is a vector function of $ x $, $ \Phi $,
and $ \boldsymbol\Lambda = (\Lamb{e}, \Lamb{i}, \Lamb{m}, \Lamb{G}) $.
The initial conditions are $ \Phi(x = 0) = (0,1,0,1) $.

The components of $ \mathfrak{F} $, as can be easily seen, are analytic functions of $  \Phi $ and $ \boldsymbol\Lambda $.
The domain of analyticity of $ \mathfrak {F} $ is a Cartesian product of the complex planes of $ \Phi $ and $ \boldsymbol \Lambda $
without $ \Phi = \infty$ and $ \boldsymbol \Lambda = \infty$.
Given that the polytropic indices are not integers,
$ \thetak{k} = 0 $ and $ \thetak{k} = \infty $ are branch points of $ \mathfrak{F} $.

According to the Poincar\'e theorem, the solutions, when they exist, depend analytically on $ \Phi (0) $ and $ \boldsymbol\Lambda $
in the domain of analyticity of $ \mathfrak {F} (x, \Phi, \boldsymbol\Lambda) $.
From Eq.~(\ref{initial_conditions}),
we observe that the dimensionless chemical potentials at $x=0$ are frozen,
so that the initial values $ \Phi(0) $ are not of interest.
The essential information is contained in $ \Lamb{e} $
and the dimensional quantities $ r_0 $ and $ \WI{n}{i0} $.
The analyticity theorem can be interpreted to mean that throughout the entire domain of their existence, solutions to Eqs.~(\ref{main_tet_1}) and (\ref{main_tet_2})
are analytic in $ \boldsymbol\Lambda$ for $ \boldsymbol\Lambda \neq \infty $.
At infinity, the solutions are either regular or irregular.

The dependence on $ \Lamb{G} $ is of particular interest because $ \Lamb{G} = \infty $ corresponds to the LEC.
It would be natural to use the LEC solution
as an initial approximation of the general solution to
Eqs.~(\ref{main_tet_1}) and (\ref{main_tet_2})
and to account for deviations from local electroneutrality
by means of perturbation theory in $ 1 / \Lamb{G} \ll 1 $.
The analyticity theorem, however, asserts that the general solution is typically not regular at $ \Lamb{G} = \infty $.
If this is so, the perturbation series in $1/\Lamb{G}$ diverges
because the radius of convergence, which is defined by the closest singularity on the complex $ \Lamb{G}$ plane, is equal to zero.

The convergence of the perturbation series for solutions that are regular at $ \Lamb{G} = \infty $
can be determined by, e.g., the singularities of $ \Lamb{e} $ and $ \Lamb{i} $ as functions of $ \Lamb{G} $,
as in the model of Sect.~II.B.
According to Eqs.~(\ref{eta eq eta}) and (\ref{ni_eq_ne_Lambe}),
$ \Lamb{e} $, $ \Lamb{i} $, and the corresponding system of ODEs
are not analytic for $ |\Lamb{G}| \sim 1 $.
If there are no singularities close to infinity,
the point $ \Lamb{G} = \infty $ lies within the circle of convergence of
the perturbation series centered at the physical value (\ref{LambdaG}).
In the model of Sect.~II.B, one could obtain then the LEC solution
in the form of a convergent series starting from the regular solution.
The reciprocal statement is also true: starting from the LEC solution,
one could obtain the regular solution in the form of a convergent series.

%%%%%%%%%%%%%%%%%%%%%%%%%%%%%%%%%%%%%%%%%%%%%%%%%%%%%%%%%%%%%%%%%%%
\subsubsection{Dyson's argument}
%%%%%%%%%%%%%%%%%%%%%%%%%%%%%%%%%%%%%%%%%%%%%%%%%%%%%%%%%%%%%%%%%%%

A physical argument leading to similar conclusions has been presented by Dyson \cite{Dyson1952}.
In an extended context, the argument %can be interpreted to
concerns the relationship
between singular points in the parameter space and
qualitative changes in the behavior of the system.
As applied to quantum electrodynamics (QED), Dyson's argument leads to the well-known
conclusions that $ \alpha = 0 $ is a singular point of physical observables, QED series in powers
of the fine structure constant have radii of convergence equal to zero and therefore are asymptotic.

In our case, $ \Lamb{G} = \infty $ corresponds to $ G = 0$.
The physical picture for $G < 0$ undergoes obvious qualitative changes; namely,
the attraction between particles is replaced by repulsion, and as a result, %for $ G <0 $,
bound states (stars) no longer exist.

We also remark that the many-body Hamilton function for $ G <0 $ is nonnegative,
whereas for positive $ G $, it is not bounded from below.

It can be observed that the characteristic length given in Eq.~(\ref{r0}) has branch points at $ G = 0 $ and $ \infty $.
The stellar mass (\ref{Mass}) and charge (\ref{Charge}) inherit these singularities,
which are further superimposed with the singularities of $ \thetak{k} $.

Dyson's argument indicates that
the general solution to Eqs.~(\ref{main_tet_1}) and (\ref{main_tet_2}) should be sought in the class of functions
that depend on $ \Lamb{G} $ in an irregular manner.

%%%%%%%%%%%%%%%%%%%%%%%%%%%%%%%%%%%%%%%%%%%%%%%%%%%%%%%%%%%%%%%%%%%
\subsection{Two-fluid model with unit polytropic indices}
%%%%%%%%%%%%%%%%%%%%%%%%%%%%%%%%%%%%%%%%%%%%%%%%%%%%%%%%%%%%%%%%%%%

The above statements can be illustrated and precisely formulated
using the model of Sect. II.B with unit polytropic indices, which admits an explicit analytic solution.
The adiabatic index of the model, $\gamma = 1 + 1/\eta = 2$, is sufficiently close to that of an ideal gas model with $\gamma = 5/3$.
In terms of the functions $ \WI{\varphi}{k} \equiv x \thetak{k} (x) $, Eqs.~(\ref{main_tet_1}) and (\ref{main_tet_2})
are reduced to a linear ODE system:
\begin{eqnarray}
\WI{\varphi}{e}^{\prime \prime} + \Lamb{i}\WI{\varphi}{i}^{\prime \prime} &=& -(\WI{\varphi}{i}+\Lamb{m}\Lamb{e}\WI{\varphi}{e}),\label{equat_simp1}\\
\WI{\varphi}{e}^{\prime \prime}-\Lamb{m}\Lamb{i}\WI{\varphi}{i}^{\prime \prime} &=& -\Lamb{G}(\WI{\varphi}{i} - \Lamb{e}\WI{\varphi}{e}),\label{equat_simp2}
\end{eqnarray}
where the prime denotes differentiation in $ x $. We look for solutions of the form $ \WI{\varphi}{k} = \WI{\alpha}{k} e^{\beta x} $.
Equations~(\ref{equat_simp1}) and (\ref{equat_simp2}) yield
\begin{eqnarray}
(\WI{\alpha}{e}+\Lamb{i}\WI{\alpha}{i})\beta^2        &=&-(\WI{\alpha}{i}+\Lamb{m}\Lamb{e}\WI{\alpha}{e}),\label{equat_alg1}\\
(\WI{\alpha}{e}-\Lamb{m}\Lamb{i}\WI{\alpha}{i})\beta^2&=&-\Lamb{G}(\WI{\alpha}{i}-\Lamb{e}\WI{\alpha}{e}).\label{equat_alg2}
\end{eqnarray}
The algebraic system formed by Eqs.~(\ref{equat_alg1}) and (\ref{equat_alg2}) has nontrivial solutions provided that the determinant is equal to zero.
This condition gives rise to a quadratic equation in $\beta^2$, whose solutions are
\begin{widetext}
\begin{eqnarray}
\beta^2_{\pm} &=& \frac{1}{2\Lamb{i}(1{+}\Lamb{m})}\bigg\{\Lamb{G}(1{+}\Lamb{i}\Lamb{e})-(1{+}\Lamb{m}^2\Lamb{i}\Lamb{e}) \label{betapm} \\
&\pm& \sqrt{\Lamb{G}^2(1{+}\Lamb{i}\Lamb{e})^2-
2\Lamb{G}\big[(1{-}\Lamb{m}\Lamb{i}\Lamb{e})^2-\Lamb{i}\Lamb{e}(1{+}\Lamb{m})^2\big]+
(1{+}\Lamb{m}^2\Lamb{i}\Lamb{e})^2}\bigg\}. \nonumber
\end{eqnarray}
\end{widetext}
To the leading order in $1/\Lamb{G} $, the values of $ \beta $ are given by
\begin{eqnarray}
\beta^2_{+}&=& \Lamb{G}\frac{1{+}\Lamb{i}\Lamb{e}}{\Lamb{i}(1 + \Lamb{m})} + O(1),  \label{equat_beta_pl}\\
\beta^2_{-}&=& -\frac{\Lamb{e}(1{+}\Lamb{m})}{1{+}\Lamb{i}\Lamb{e}}+ O(\Lamb{G}^{-1}).  \label{equat_beta_mn}
\end{eqnarray}
The values $\pm \sqrt{\beta^2_{+}} $ are real and large, whereas the values $\pm \sqrt{\beta^2_{-}}$ are imaginary.
We adopt the notation $ \beta_1 = |\beta_{+}| > 0$ and $ \beta_2 = |\beta_{-} |> 0 $.
The general solution has the form
\begin{eqnarray}
\WI{\varphi}{k}(x)&=&\WI{\widetilde{\alpha}}{k1}\exp(-\beta_1 x)+\WI{\widetilde{\alpha}}{k2}\exp(\beta_1 x)\nonumber \\
&+&\WI{\widetilde{\alpha}}{k3}\sin(\beta_2 x)+\WI{\widetilde{\alpha}}{k4}\cos(\beta_2 x).\label{phi_general}
\end{eqnarray}
The initial conditions for $ \thetak{k} $ lead to
$ \WI{\varphi}{k} (0) = 0 $, $ \WI{\varphi}{k}^{\prime}(0) = 1 $, and $ \WI{\varphi}{k}^{\prime \prime}(0) = 0 $.
The general solution becomes
\begin{equation}
\WI{\varphi}{k}(x)=\WI{\alpha}{k1}\sinh(\beta_1 x)+\WI{\alpha}{k2}\sin(\beta_2 x),\label{phi_general2}
\end{equation}
with the constraint
\begin{equation}
\WI{\alpha}{k1}\beta_1+\WI{\alpha}{k2}\beta_2=1.\label{cond_alpha}
\end{equation}
By substituting $\beta^2 = \beta_1^2$ and $- \beta_2^2$ into Eq.~(\ref{equat_alg1}) for $\WI{\alpha}{k1}$ and $\WI{\alpha}{k2}$, respectively,
and using Eq.~(\ref{cond_alpha}), we can uniquely determine $\WI{\alpha}{k1}$ and $\WI{\alpha}{k2}$ and find the solution.

%%%%%%%%%%%%%%%%%%%%%%%%%%%%%%%%%%%%%%%%%%%%%%%%%%%%%%%%%%%%%%%%%%%
\subsubsection{Closer inspection of analyticity in $G$}
%%%%%%%%%%%%%%%%%%%%%%%%%%%%%%%%%%%%%%%%%%%%%%%%%%%%%%%%%%%%%%%%%%%

The analytic properties of $ \WI{\varphi}{k} (x) $ in the complex $ \Lamb{G} $-plane can be clarified by
transforming the system defined by Eqs.~(\ref{equat_simp1}) and (\ref{equat_simp2}) into normal form with the
independent variables $ \WI{\varphi}{k} $ and $ \WI{\pi}{k} = \WI{\varphi}{k}^{\prime} $.
In terms of $ \Phi \equiv ( \WI{\pi}{i}, \WI{\varphi}{i}, \WI{\pi}{e}, \WI{\varphi}{e} )$,
Eqs.~(\ref{equat_simp1}) and (\ref{equat_simp2})
take the form
\begin{equation}
\Phi^{\prime} = \mathfrak {A} \Phi,
\end{equation}
where $ \mathfrak {A} $ is a real $ 4 \times 4 $ matrix that
depends linearly on the parameters
$ \boldsymbol\Lambda = (\Lamb{e}, \Lamb{i}, \Lamb{m}, \Lamb{G}) $. The solution takes the form
\begin{equation} \label{inidata}
\Phi (x) = \exp (\mathfrak {A} x) \Phi (0).
\end{equation}
The series expansion of the exponential matrix converges,
so the evolution operator $ \mathfrak {D} = \exp (\mathfrak {A} x) $ is well defined.
The matrix elements of $ \mathfrak {D} $ are analytic functions in $ \boldsymbol\Lambda $ for $ \boldsymbol\Lambda \neq \infty$.
A simple pole of $ \mathfrak {A} $ at $ \boldsymbol\Lambda = \infty $ is transformed
into an essential fixed singularity of $ \mathfrak {D} $, which does not depend on the initial values.
The functions $ \Phi (x) $ are, therefore, analytic functions
of $ \boldsymbol \Lambda $ in a Cartesian product of the complex planes of $\Lamb{e}$, $\Lamb{i}$, $\Lamb{m}$, and $\Lamb{G}$.
Equation (\ref{inidata}) also shows that $ \Phi (x) $ are analytic on the initial values $ \Phi (0) $,
in agreement with the Poincar\'e theorem on analyticity.

Let $ \mathfrak{a}_{n} $ denote the eigenvalues of $ \mathfrak{A} $ that coincide with $ \pm \beta_{\pm} \in \mathbb{C}^1 $,
and let $ \mathfrak{r}_{n} $ and $ \mathfrak{l}_{n} $ be their corresponding right and left eigenvectors
satisfying $ \mathfrak{A} \mathfrak {r}_{n} = \mathfrak{a}_{n} \mathfrak{r}_{n} $ and
$ \mathfrak{l}_{n}^{T } \mathfrak {A} = \mathfrak {l}_{n}^{T} \mathfrak {a}_{n} $. The eigenvectors are normalized according to
$ \mathfrak {l}_{n}^{T} \mathfrak {r}_{m} = \delta_{nm} $, where $ n = 1,2,3,4 $.
The evolution operator admits the representation
\begin{equation}
\mathfrak {D} = \sum_{n} \mathfrak {r}_{n} \mathfrak {l}_{n}^{T} \exp (\mathfrak {a}_{n} x).
\end{equation}
The eigenvalues $ \mathfrak{a}_{n} $ are roots of the characteristic equation,
so they have additional singularities compared with $ \mathfrak {A} $.
The analyticity of $\mathfrak {D}$ implies that additional singularities cancel out with
singularities of the projection operators $ \mathfrak {r}_{n} \mathfrak {l}_{n}^{T} $.

A more explicit proof of this claim uses the Frobenius representation
\begin{equation} \label{Frobenius}
\mathfrak {r}_{n} \mathfrak {l}_{n}^{T} = \prod_{m \neq n} \frac {\mathfrak {A} - \mathfrak {a}_{m}} {\mathfrak {a}_{n} - \mathfrak {a}_{m}}.
\end{equation}
We first note that the radicand in Eq.~(\ref{betapm}) vanishes for a complex-conjugate
pair of $ \Lamb{G} $, where $ \beta_{\pm} $ have square-root branch points.
The point $ \Lamb{G} = \infty $ is a simple pole of $ \beta_{+}^2 $,
a square-root branch point of $ \beta_{+} $, and an essentially singular point of $ \varphi_{\mathrm {k}} (x) $.
$ \Lamb{G} = 0 $ is a simple zero of $ \beta_{+}^2 $ and a square-root branch point of
$ \beta_{+} $.
The eigenvalues $ \mathfrak {a}_{n} $ ($\pm \beta_{\pm}$) accordingly have four square-root branch points in the complex $ \Lamb{G} $ plane.
By passing around one of these points, we obtain simply a permutation of $ \mathfrak {a}_{n} $.
The representation of $ \mathfrak {D} $ using Eq.~(\ref{Frobenius}) is explicitly invariant under these permutations,
so the matrix elements of $ \mathfrak {D} $ do not
inherit additional singularities of $ \mathfrak {a}_{n} $.

Consequently, the eigenvalues $\pm \beta_{\pm}$ and the projection operators of $ \mathfrak{A} $ have four singular points,
whereas $ \varphi_{\mathrm {k}} (x) $ has a unique singular point $ \Lamb{G} = \infty $.

Given that $ \WI{\varphi}{k}(x) $ is analytic in $ \Lamb{G} \in \mathbb{C}^1 \setminus \infty $,
$ \WI{\varphi}{k}(x) $ if not a constant must be singular at $ \Lamb{G} = \infty $. %, otherwise $\varphi_{k}(x) = \mathrm{const}$.
The existence of solutions independent of $ \Lamb{G} $ is excluded by Eq.~(\ref{equat_simp2}).
Regular solutions at $ \Lamb{G} = \infty $ can be constructed by replacing the parameters, e.g., $ \Lamb{e} $ and $ \Lamb{i} $,
with functions of $\Lamb{G}$, as in the model of Sect.~II.B.
Solutions inherit the singularities of these functions,
which makes it possible to construct a regular solution.
The dependence of the parameters on $ \Lamb{G} $ is a necessary %(but not sufficient)
condition for the existence of a regular solution in the neighborhood of $ \Lamb{G} = \infty $.

A qualitative analysis of Eqs.~(\ref{equat_simp1}) and (\ref{equat_simp2}), developed above, reveals
the mathematical reason why the parameters $ \Lamb{e} $ and $ \Lamb{i} $
are not independent in the regular solution.
The analyticity in $ \Lamb{G} $ follows from the representation given in Eq.~(\ref{inidata}).
The requirement of regularity at $ \Lamb{G} = \infty $
would imply regularity in the extended complex $ \Lamb{G}$ plane.
In such a case, $ \WI{\varphi}{k}(x) $ must be a constant.
By introducing a dependence of the parameters on $ \Lamb{G} $,
particular, non-trivial, regular solutions at $ \Lamb{G} = \infty $ can be constructed,
which can further be used to approximate the general solution.

%%%%%%%%%%%%%%%%%%%%%%%%%%%%%%%%%%%%%%%%%%%%%%%%%%%%%%%%%%
\subsubsection{Electro- and ionospheres}
%%%%%%%%%%%%%%%%%%%%%%%%%%%%%%%%%%%%%%%%%%%%%%%%%%%%%%%%%%

The first term in Eq.~(\ref{phi_general2}) depends exponentially on $ \beta_1 \sim \sqrt{\Lamb{G}} $.
For $ x \neq 0 $, $\sinh (\beta_1 x) $ has no zeros. For the chemical potential to vanish at the boundary,
the solution must contain the component $ \sim \sin (\beta_2 x) $. The first term in Eq.~(\ref{phi_general2})
is maximal on the boundary of the star, where it should not exceed $ \alpha_2 $. The ratio $ \alpha_1 / \alpha_2$
is therefore exponentially small.

The solutions are recovered with exponential accuracy already for $ \alpha_2 \beta_2 = 1 $.
Substituting this expression into Eqs.~(\ref{equat_alg1}) and (\ref{equat_alg2}), we obtain
\begin{eqnarray}
(1{+}\Lamb{i})\beta_2^2        &=&1{+}\Lamb{m}\Lamb{e},\label{equat_alg1_beta}\\
(1{-}\Lamb{m}\Lamb{i})\beta_2^2&=&\Lamb{G}(1{-}\Lamb{e}).\label{equat_alg2_beta}
\end{eqnarray}
The self-consistency of the system leads to $ \Lamb{e} = \Lamb{e}^{\mathrm{reg}} $,
where $\Lamb{e}^{\mathrm{reg}}$ is given by Eq.~(\ref{ni_eq_ne_Lambe}).
The corresponding $ \beta $ parameters (with an additional index 0) become
\begin{eqnarray}
\beta_{10}^2&=&\frac{\Lamb{G}(1{+}\Lamb{i})-(1{-}\Lamb{m}\Lamb{i})}{\Lamb{i}(1{+}\Lamb{m})},\label{beta_10}\\
\beta_{20}^2&=&\frac{\Lamb{G}(1{+}\Lamb{m})}{\Lamb{G}(1{+}\Lamb{i})+\Lamb{m}(1{-}\Lamb{m}\Lamb{i})}.\label{beta_20}
\end{eqnarray}
Equations (\ref{ni_eq_ne_Lambe}), (\ref{beta_10}) and (\ref{beta_20}) are exact to all orders in $ 1 / \Lamb{G} $.
$ \Lamb{i} $ is a function of $ \Lamb{G} $, according to Eq.~(\ref{eta eq eta}).
In the approximation $ \alpha_1=0 $, with exponential accuracy, the function $ \theta (x) $ is given by
\begin{equation}
\theta(x)=\frac{\varphi(x)}{x}\approx\frac{\sin(\beta_{20} x)}{\beta_{20} x},\label{eta1_classical}
\end{equation}
in agreement with the analysis of Sect.~II.B for unit polytropic indices.
The function $ \theta (x) $ is regular for $ \Lamb{G} = \infty$ but inherits the singularities of $\beta_ {20}$.

A variation in the electron density at $r=0$ leads to a variation $\triangle \Lamb{e}  = \Lamb{e} - \Lamb{e}^{\mathrm{reg}} \neq 0$,
corresponding variations of $ \Lamb{i} $ and $ \beta_{\pm} $ and, ultimately, to $\WI{\alpha}{k 1} \neq 0$.
We return to the general solution (\ref{phi_general2}) and write the function $ \thetak{k} $ in the form
\begin{equation}
\thetak{k}(x)=\WI{\alpha}{k 1} \frac{\sinh(\beta_1 x)}{x} + \WI{\alpha}{k 2} \frac{\sin(\beta_2 x)}{x},\label{theta_ie}
\end{equation}
where, with account of Eq.~(\ref{cond_alpha}),
\begin{equation} \label{alpha2}
\WI{\alpha}{k 2} = \frac{1-\WI{\alpha}{k1}\beta_1}{\beta_2}.
\end{equation}
The regular solution (\ref{eta1_classical}) vanishes for $ x_0 = \pi / \beta_ {20} $.
For $ \WI{\alpha}{k1} \neq 0 $, one of the functions $ \thetak{k} (x) $ vanishes
already at $ x_ {b} = x_0 + \triangle x $, whereas the second function is still finite.
We write this condition in the form
\begin{eqnarray}
\thetak{i}(x_{b})&=&0,\label{theta_i_0} \\
\thetak{e}(x_{b})&=&\thetak{eb},\label{theta_e_0}
\end{eqnarray}
where $\thetak{eb}$ is the value of the electron function at the boundary. We assume first that an electrosphere is formed.

As follows from Sect.~III, the electron pressure at the boundary of the star is $ \WI{P}{eb} \sim \WI{P}{un} / \Lamb{G} $,
where $ \WI{P}{un} $ is the characteristic pressure scale. The relation $ \WI{P}{e} \sim \WI{n}{e}^{1 + 1/ \etak{e}} \sim \thetak{e}^{1+ \etak{e}} $
implies that $ \thetak{eb} \sim \Lamb{G}^{- 1 / (1 + \etak{e})} \ll 1 $. For the case under consideration, $ \etak{e} = 1 $, and accordingly,
$ \thetak{eb} \sim 1 / \sqrt {\Lamb{G}} $. The typical scale of the electrosphere is $\WI{r}{a} / r_{0} \sim \thetak{eb} \sim \Lamb{G}^{-1 / (1 + \etak{e})} $,
in agreement with the estimates given in Eqs.~(\ref{r0_non_rel}) and (\ref{r0_rel}). Using Eq.~(\ref{gradP}) and the inequality $ W_0 \geq 0 $, we obtain
\begin{widetext}
\begin{equation} \label{atmomax}
\thetak{eb} \leq \WI{\lambda}{G}^{-1/(1+\etak{e})}
\left[ \frac{\left( 1+\etak{e}\right) (\WI{q}{s}+\WI{\lambda}{m})^{2}}
{2\Lamb{e}(1+\Lamb{m})(1-\WI{\lambda}{m}^{2}/\WI{\lambda}{G})}
\left( \frac{\WI{M}{s}}{%
4\pi r_{0}^{3} \WI{n}{i0}\WI{m}{i}} \right) ^{2}\left( \frac{%
r_{0}}{\WI{R}{s}}\right) ^{4}\right] ^{1/(1+\etak{e})}.
\end{equation}
\end{widetext}

Equations (\ref{theta_i_0}) and (\ref{theta_e_0}) allow to express the unknown coefficients $ \WI{\alpha}{k 1} $ in terms of $ \thetak{eb} $.
Given the relation in Eq.~(\ref{alpha2}), the functions $ \thetak{k} (x) $ are completely defined, but the parameters $ \triangle x $ and $ \thetak{eb} $
on which these functions depend must be found. We first substitute $ \WI{\alpha}{k 1} $ into the first of Eqs.~(\ref{equat_alg1}) and (\ref{equat_alg2}), setting $ \beta^2 = \beta_1^2 $, and then substitute $ \WI{\alpha}{k 2} $, setting $ \beta^2 = - \beta_2^2 $. The two resulting equations are linear in $ \thetak{eb} $. By excluding $ \thetak{eb} $, we find
\begin{eqnarray} \label{dxvsdl}
&\;&{\frac {
(1+\Lamb{i})\beta_2^2 - 1 - \Lamb{m}\Lamb{e} }{\beta_2^2  - \Lamb{m}\Lamb{e}}} \frac{\sinh \left( \beta_1 \WI{x}{b} \right)}{\beta_1 \WI{x}{b} } \nonumber \\
 &=&
\frac{ (1 + \Lamb{i})\beta_1^2 +1+\Lamb{m}\Lamb{e}}
{ \beta_1^2 + \Lamb{m}\Lamb{e}} \frac{\sin( \beta_2 \WI{x}{b} )}{\beta_2 \WI{x}{b} }.
\end{eqnarray}
This equation defines $ \WI{x}{b} $ as a function of the parameters $ \Lamb{e} $, $ \Lamb{i} $, $ \Lamb{m} $, and $ \Lamb{G} $.
Equation (\ref{equat_alg2}) is not used because for $ \beta^2 = \beta_{\pm}^2 $, it is not independent. Both parts of Eq.~(\ref{dxvsdl})
define $ \thetak{eb} \geq 0 $; therefore, a solution can exist for $ \beta_2^2 \geq \beta_{20}^2 $ and $ \triangle x \leq 0 $.
As noted above, $ \thetak{eb} \sim 1 /\beta_{1} \sim 1 / \sqrt {\Lamb{G}} $, and so
$ \triangle x \sim \WI{r}{a} / r_{0} \sim 1 / \sqrt {\Lamb{G}} $.
Considering $ \triangle x \sim 1 / \beta_ {1} \ll 1 $, the right-hand side of Eq.~(\ref{dxvsdl}) can be expanded into a series
in the neighborhood of $ \WI{x}{b} = {x}_{0} $; as a result, we find
\begin{equation} \label{delta x}
\triangle x = - \frac{1}{\beta_{1}}\mathcal{W}\left( \frac{\triangle\Lamb{e}\exp\left({\pi\beta_{1}}/{\beta_{2}}\right)}{2(1 + \Lamb{i})}\right).
\end{equation}
The displacement $ \triangle x $ is expressed in terms of the Lambert $W$-function, which gives a solution to the equation $ \mathcal{W} (x) \exp (\mathcal {W} (x)) = x $. Since the argument of the function is of order unity, the correction to $ \Lamb{e} $ is exponentially small:
\[
\triangle\Lamb{e}\sim \exp\left(-O(\sqrt{\Lamb{G}})\right).
\]
For $ \thetak{eb} $, we obtain
\begin{equation}
\thetak{eb} = -  \frac{\triangle x }{x_{0}}(1 + \Lamb{i}).
\label{theta_eb}
\end{equation}
In Eqs.~(\ref{delta x}) and (\ref{theta_eb}), the $ \sim \exp \left (- \pi \beta_ {1} / \beta_{2} \right) $ terms and
higher terms in $ 1 / \sqrt{\Lamb{G}} $ are omitted. At this level of accuracy, we can replace $ \beta_{1} $ and $ \beta_{2} $
with the expressions in Eqs.~(\ref{beta_10}) and (\ref{beta_20}) everywhere, including exponents. The positive definiteness of $ \thetak{eb} $ implies that
$ \triangle x <0 $ and $ \triangle \Lamb{e} > 0 $.

For asymptotically large values of the argument of $ \mathcal {W} (x) $, the expansion $ \mathcal{W} (x) = \ln (x) - \ln (\ln (x)) + \ldots $ is valid.
A correction $ \triangle \Lamb{e} \sim 1 / \sqrt {\Lamb{G}} $ could lead
to the estimate $ \triangle x \sim \thetak{eb} \sim 1 $, which contradicts the upper limit of Eq.~(\ref{atmomax}).
The set of admissible solutions, therefore, consists only of solutions with exponentially small deviations $ \triangle \Lamb{e} $.

Equations (\ref{equat_simp1}) and (\ref{equat_simp2}) also permit solutions corresponding to an ionosphere.
Instead of Eqs.~(\ref{theta_i_0}) and (\ref{theta_e_0}), we require
\begin{eqnarray}
\thetak{i}(x_{b})&=&\thetak{ib}, \label{theta_i_0 bis} \\
\thetak{e}(x_{b})&=&0. \label{theta_e_0 bis}
\end{eqnarray}
The equation for determining $ \triangle x $ is as follows:
\begin{eqnarray} \label{dxvsd2}
&\;&\frac
{(1+\Lamb{i})\beta_2^2 - 1 - \Lamb{m}\Lamb{e} }
{\beta_2^2 \Lamb{i} - 1} \frac{\sinh \left( \beta_1 \WI{x}{b} \right)}{\beta_1 \WI{x}{b} } \nonumber \\
&=&
\frac{ (1 + \Lamb{i})\beta_1^2 +1+\Lamb{m}\Lamb{e}}
{ \beta_1^2 \Lamb{i} + 1} \frac{\sin( \beta_2 \WI{x}{b} )}{\beta_2 \WI{x}{b} }.
\end{eqnarray}
Both parts are equal to $ \thetak{ib}> 0 $. The relationship between $ \triangle x $, $ \triangle \Lamb{e} $ and $ \thetak{ib} $
is established by the relations (\ref{delta x}) and (\ref{theta_eb}) with the substitutions
$ \thetak{eb} \to \thetak{ib} $, $ \triangle \Lamb{e} \to - \triangle \Lamb{e} $ and $ \Lamb{i} \to 1 / \Lamb{i} $, while the other parameters remain unchanged.
The condition $ \thetak{ib}> 0 $ implies that $ \triangle x <0 $ and $ \triangle \Lamb{e} <0 $.
An upper limit on $ \thetak{ib} $ is found from the inequality $ W_0 \geq 0 $
and the balance between the pressure and the forces of gravitational attraction and electrostatic repulsion acting on the ions at the boundary of the electron component (cf. (\ref{gradP})):
\begin{widetext}
\begin{equation} \label{atmoimax}
\thetak{ib} \leq \WI{\lambda}{G}^{-1/(1+\etak{i})}
\left[ \frac{(1+\etak{i})(- \WI{q}{s}+ \WI{\lambda}{m}/\Lamb{m})^{2}}
{2\Lamb{i}(1+\Lamb{m})(1-1/\Lamb{G})}
\left( \frac{\WI{M}{s}}{4\pi r_{0}^{3} \WI{n}{i0} \WI{m}{i}}\right) ^{2}\left( \frac{r_{0}}{\WI{R}{s}
}\right) ^{4}\right] ^{1/(1+\etak{i})}.
\end{equation}
\end{widetext}
An ionosphere exists provided that $ \WI{q}{s} < \WI{\lambda}{m} / \Lamb{m} = A / Z $.

For positive values of $ \triangle \Lamb{e} $, the substance in the center of the star
is closer to the electrically neutral state than in the regular solution.
On the surface of the baryon component, the bulk charge is positive, and the electrosphere compensates for this charge either partially, completely, or excessively.
Suppression of the electron fraction increases the deviation from electroneutrality and corresponds to $ \triangle \Lamb{e} < 0$.
Under the condition $ \WI{q}{s} < A / Z $, the electron density vanishes when the ion density is still finite.
On the surface of the electron component, the stellar charge is positive (by continuity) and increases with $ | \triangle \Lamb{e} | $.
The ionosphere makes an additional contribution to this charge.
$ | \triangle \Lamb{e} | $ can be increased until it violates (\ref{atmomax}) in the case of an electrosphere or
(\ref{atmoimax}) in the case of an ionosphere.

%%%%%%%%%%%%%%%%%%%%%%%%%%%%%%%%%%%%%%%%%%%%%%%%%%%%%%%%%%
\subsubsection{Charge-mass-radius relation}
%%%%%%%%%%%%%%%%%%%%%%%%%%%%%%%%%%%%%%%%%%%%%%%%%%%%%%%%%%

The general properties associated with a deviation from local electroneutrality are expected to be inherent in realistic models. We thus consider the charge-mass-radius relation for stars. The mass and radius are expressed as follows:
\begin{eqnarray}
\frac{\WI{M}{s}}{4\pi r_{0}^3 n_{\mathrm{i}0}m_{\mathrm{i}}} &=& \frac{\pi (1 + \Lamb{i})}{\beta_2}, \label{mass 12} \\
\frac{\WI{R}{s}}{r_{0}} &=& \frac{\pi}{\beta_2} \label{radius 12}.
\end{eqnarray}
The violation of the LEC affects the mass and radius at higher orders of the expansion in $ 1 / \sqrt {\Lamb{G}} $.
The correction to the radius is $ \Delta \WI{R}{s} = r_0 \triangle x \sim r_0 / \sqrt {\Lamb{G}} $,
so the mass-radius relation remains unchanged to within $ \sim 1 / \sqrt{\Lamb{G}} $.
The bulk charge of the star, $ \WI{q}{s} $, and the electrosphere charge, $ \WI{q}{e} $, are as follows:
\begin{widetext}
\begin{eqnarray}
\frac{Z}{A}\WI{q}{s} &=&\frac{1-\Lamb{m}\Lamb{i}}{1+\Lamb{i}}-\frac{%
(1+\Lamb{m})\Lamb{i}}{(1+\Lamb{i})^{2}}\pi \WI{\Theta}{eb}, \label{charge 1} \\
\frac{Z}{A} \WI{q}{e} &=& -\frac{1+\Lamb{m}}{1+\Lamb{i}}\left[ -\sqrt{\left( 1-\frac{\Lamb{i}}{1+\Lamb{i}}\pi
\WI{\Theta}{eb}\right) ^{2}-\frac{\Lamb{i}}{%
1+\Lamb{i}}\pi ^{2}\WI{\Theta}{eb}^{2}}+\left( 1-\frac{\Lamb{i}}{%
1+\Lamb{i}}\pi \WI{\Theta}{eb}\right) \right], \nonumber
\end{eqnarray}
\end{widetext}
where $ \WI{\Theta}{eb} = \beta_1 \thetak{eb} / \beta_2 \sim $ 1.
Both formulae are valid with an accuracy of $ \sim 1 / \sqrt {\Lamb{G}} $. The existence of an electron shell
requires an excess of electrons relative to the solution $ \thetak{eb} = \thetak{ib} = 0 $.
Accordingly, the second term in $ \WI{q}{s} $ is negative, and as $ \WI{\Theta}{eb} $ increases, its contribution reduces
the stellar charge. By equating the root expression to zero, we can obtain the maximum value of $ \Theta_{eb} $:
\begin{equation}
\WI{\Theta}{eb}^{\max} = \frac{(\sqrt{1 + 1/\Lamb{i} } - 1)(1 + \Lamb{i})}{\pi}.
\end{equation}
This value corresponds to the upper limit in Eq.~(\ref{atmomax}). For $ \WI{\Theta}{eb} = \WI{\Theta}{eb}^{\max} $, the shell overcompensates
for the positive charge of the star: $ (\WI{q}{s} + \WI{q}{e})^{\min} = - \WI{\lambda}{m} <0 $.

An ionosphere forms for a lower concentration of electrons.
The bulk charge, $ \WI{q}{s} $, measured at the boundary of the electron component and the ion envelope charge, $ \WI{q}{e} $, are given by the expressions
\begin{widetext}
\begin{eqnarray}
\frac{Z}{A}\WI{q}{s} &=&\frac{1-\Lambda _{\mathrm{m}}\Lambda _{\mathrm{i}}}{1+\Lambda _{\mathrm{i}}}
+\frac{(1+\Lambda _{\mathrm{m}})\Lambda _{\mathrm{i}}}{(1+\Lambda _{\mathrm{i}})^{2}}\pi\WI{\Theta}{ib}, \label{charge 2} \\
\frac{Z}{A} \WI{q}{e} &=& \frac{(1+\Lambda _{\mathrm{m}})\Lambda _{\mathrm{i}}}{1+\Lambda _{\mathrm{i}}}
\left[ -\sqrt{\left( 1-\frac{1}{1+\Lambda _{\mathrm{i}}}\pi\WI{\Theta}{ib}\right) ^{2}-\frac{1}{%
1+\Lambda _{\mathrm{i}}}\pi ^{2}\Theta _{\mathrm{ib}}^{2}}+\left( 1-\frac{1}{%
1+\Lambda _{\mathrm{i}}}\pi\WI{\Theta}{ib}\right)\right], \nonumber
\end{eqnarray}
\end{widetext}
where $ \WI{\Theta}{ib} = \beta_1 \thetak{ib} / \beta_2 \sim 1$.
The charge $ \WI{q}{s} $ increases with $ \WI{\Theta}{ib} $. The radicand vanishes
when $ \WI{\Theta}{ib} $ takes its maximum value
\begin{equation}
\WI{\Theta}{ib}^{\max} = \frac{(\sqrt{1 + \Lamb{i}} - 1)(1 + 1/\Lamb{i})}{\pi}.
\label{limit 2}
\end{equation}
One can check that for $ \WI{\Theta}{ib} = \WI{\Theta}{ib}^{\max} $, the inequality in Eq.~(\ref{atmoimax})
becomes an equality. The maximum total charge is equal to $ (\WI{q}{s} + \WI{q}{e})^{\max} = A / Z $.
The condition $ \WI{q}{s} < A / Z $ constrains the value of $ \WI{\Theta}{ib} $ from above by
$ (1 + \Lamb{i}) / \pi $. However, this limit is weaker than $ \WI{\Theta}{ib} \leq \WI{\Theta}{ib}^{\max} $.

\begin{figure} [t] %
\begin{center}
\includegraphics[angle = 0,width=0.45\textwidth]{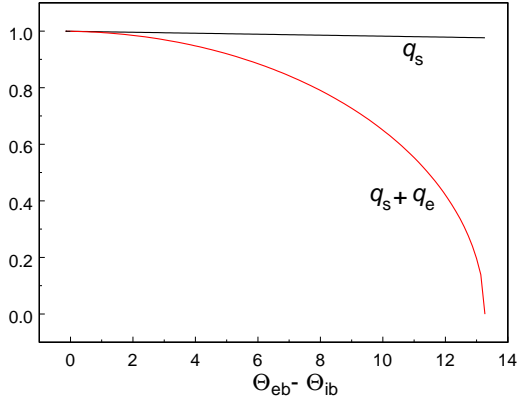}
\vspace{3 mm}
\caption{(color online)
The bulk charge $ \WI{q}{s} $ and the total charge $ \WI{q}{s} + \WI{q}{e} $ as functions of the difference $ \WI{\Theta}{eb} - \WI{\Theta}{ib} $
for a mixture of electrons and protons ($ Z = A = 1 $) and for $ \Lamb{i} = \Lamb{m} $.
Positive values of $ \WI{\Theta}{eb} - \WI{\Theta}{ib} $ correspond to the formation of an electrosphere ($ \WI{\Theta}{ib} = 0 $),
whereas negative values correspond to the formation of an ionosphere (proton envelope, $ \WI{\Theta}{eb} = 0 $).
The region near zero is shown on an enlarged scale in Fig. \ref{fig4}.
}
\label{fig3}
\end{center}
\end{figure}

\begin{figure} [b] %
\begin{center}
\includegraphics[angle = 0,width=0.45\textwidth]{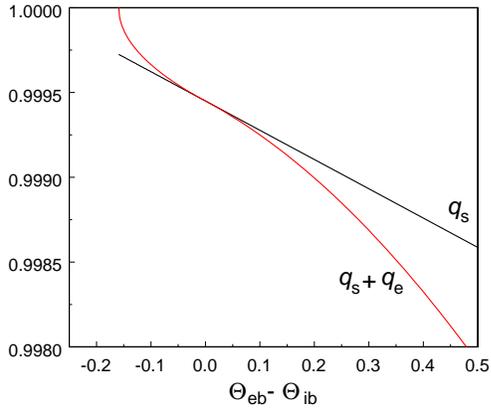}
\vspace{3 mm}
\caption{(color online) The bulk charge $ \WI{q}{s} $ and the total charge $ \WI{q}{s} + \WI{q}{e} $
as functions of the difference $ \WI{\Theta}{eb} - \WI{\Theta}{ib} $ on an enlarged scale. The notation is the same as in Fig.~\ref{fig3}.
}
\label{fig4}
\end{center}
\end{figure}

Figure \ref{fig3} plots the bulk charge of the star, $ \WI{q}{s} $, and the total charge including the envelope, $ \WI{q}{s} + \WI{q}{e} $,
as functions of the difference $ \WI{\Theta}{eb} - \WI{\Theta}{ib} $
in the interval $ (- \WI{\Theta}{ib}^{\max}, \WI{\Theta}{eb}^{\max}) $ for a mixture of electrons and protons, i.e., for
$ Z = A = 1 $ and $ \Lamb{i} = \Lamb{m} $. The region $ \WI{\Theta}{eb} - \WI{\Theta}{ib} \sim 0$ is presented separately in Fig.~\ref{fig4} on an enlarged scale.

In summary, in the two-fluid model with unit polytropic indices,
it is possible to explicitly construct the general solution describing charged stars with electro- and ionospheres.
An electrosphere forms if the ion chemical potential vanishes at the boundary, $ \thetak{i} (\WI{x}{b}) = 0 $,
whereas the electron chemical potential at the boundary is finite, $ \thetak{e} (\WI{x}{b})> 0 $, according to Eqs.~(\ref{theta_i_0}) and (\ref{theta_e_0}).
An ionosphere forms in the opposite case; see Eqs.~(\ref{theta_i_0 bis}) and (\ref{theta_e_0 bis}).
The solution for the internal stellar structure is unambiguously matched with the solution for the electro- or ionosphere.
The guiding parameter of the problem is $ \triangle \Lamb{e} = \Lamb{e} - \Lamb{e}^{\mathrm{reg}}$.
The general solution also describes stars with zero total charge, but such stars are not locally neutral.
In an exponentially small neighborhood of $ \Lamb{e}^{\mathrm{reg}} $, the electron and ion densities
can be varied without restrictions. The stellar envelope is sensitive to {exponentially} small deviations of $ \Lamb{e} $
from $ \Lamb{e}^{\mathrm{reg}} $.
The same fact can be interpreted as indicating remarkable overall stellar stability: changes in the envelope, and, accordingly, in the total stellar charge (within acceptable limits) have an exponentially weak influence on the internal structure of a star.

%%%%%%%%%%%%%%%%%%%%%%%%%%%%%%%%%%%%%%%%%%%%%%%%%%%%%%%%%%%%%%%%%%%%%%%%%%%%
%%%%%%%%%%%%%%%%%%%%%%%%%%%%%%%%%%%%%%%%%%%%%%%%%%%%%%%%%%%%%%%%%%%%%%%%%%%%
%%%%%%%%%%%%%%%%%%%%%%%%%%%%%%%%%%%%%%%%%%%%%%%%%%%%%%%%%%%%%%%%%%%%%%%%%%%%
%%%%%%%%%%%%%%%%%%%%%%%%%%%%%%%%%%%%%%%%%%%%%%%%%%%%%%%%%%%%%%%%%%%%%%%%%%%%
%%%%%%%%%%%%%%%%%%%%%%%%%%%%%%%%%%%%%%%%%%%%%%%%%%%%%%%%%%%%%%%%%%%%%%%%%%%%
%%%%%%%%%%%%%%%%%%%%%%%%%%%%%%%%%%%%%%%%%%%%%%%%%%%%%%%%%%%%%%%%%%%%%%%%%%%%

%%%%%%%%%%%%%%%%%%%%%%%%%%%%%%%%%%%%%%%%%%%%%%%%%%%%%%%%%%%%%%%%%%%%%%%%%%%%
%%%%%%%%%%%%%%%%%%%%%%%%%%%%%%%%%%%%%%%%%%%%%%%%%%%%%%%%%%%%%%%%%%%%%%%%%%%%
%%%%%%%%%%%%%%%%%%%%%%%%%%%%%%%%%%%%%%%%%%%%%%%%%%%%%%%%%%%%%%%%%%%%%%%%%%%%
%\vspace{-5cm}
%%%%%%%%%%%%%%%%%%%%%%%%%%%%%%%%%%%%%%%%%%%%%%%%%%%%%%%%%%%%%%%%%%%%%%%%%%%%
%%%%%%%%%%%%%%%%%%%%%%%%%%%%%%%%%%%%%%%%%%%%%%%%%%%%%%%%%%%%%%%%%%%%%%%%%%%%
%%%%%%%%%%%%%%%%%%%%%%%%%%%%%%%%%%%%%%%%%%%%%%%%%%%%%%%%%%%%%%%%%%%%%%%%%%%%

%%%%%%%%%%%%%%%%%%%%%%%%%%%%%%%%%%%%%%%%%%%%%%%%%%%%%%%%%%%%%%%%%%%%%%%%%%%%
%%%%%%%%%%%%%%%%%%%%%%%%%%%%%%%%%%%%%%%%%%%%%%%%%%%%%%%%%%%%%%%%%
\section{General solution to the unconstrained hydrostatic equilibrium equations}
%%%%%%%%%%%%%%%%%%%%%%%%%%%%%%%%%%%%%%%%%%%%%%%%%%%%%%%%%%%%%%%%%
%%%%%%%%%%%%%%%%%%%%%%%%%%%%%%%%%%%%%%%%%%%%%%%%%%%%%%%%%%%%%%%%%%%%%%%%%%%%
\setcounter{equation}{0}

From the analysis of the exactly solvable model, the specific nature of the problem becomes rather transparent:
the general solution to Eqs.~(\ref{main_tet_1}) and (\ref{main_tet_2}) contains a component that is irregular at $ \Lamb{G} = \infty $.
In the model of Sect.~IV, the regular component dominates up to the surface of the star,
whereas the irregular component is exponentially small everywhere except for the subsurface layer
$ \Delta r \sim \WI{r}{a} \sim r_0 /\sqrt{\Lamb{G}}$, where
the contributions of the two components are comparable. The exponential smallness of the irregular component makes perturbation
theory appropriate for describing the inner layers of the star.

There are four independent parameters, $ \Lamb{e} $, $ \Lamb{i} $, $ \Lamb{m} $, and $ \Lamb{G} $,
on which Eqs.~(\ref{main_tet_1}) and (\ref{main_tet_2}) and their solutions depend.
The regular solution is a three-parameter solution with the independent parameters $ \Lamb{i} $, $ \Lamb{m} $, and $ \Lamb{G} $.
The function $ \Lamb{e} = \Lamb{e}^{\mathrm{reg}} $ is defined in Sect.~II.C to $ O(1/\Lamb{G} )$.
The general solution and the regular solution satisfy the equations with different values of $ \Lamb{e} $.
We denote the variance by $ \triangle \Lamb{e} = \Lamb{e} - \Lamb{e}^{\mathrm{reg}}$, $|\triangle \Lamb{e}|  \ll 1 $.
The parameter $ \Lamb{i} $ is not modified since for the general case $ \etak{e} \ne \etak{i} $ considered here,
$ \Lamb{e} $ and $ \Lamb{i} $ are independent.

Let $ \thetak{k0} $ denote the regular solution. The general solution is assumed to be of the form
\begin{equation} \label{general}
\thetak{k} = \thetak{k0} + \WI{\chi}{k},
\end{equation}
where $ \WI{\chi}{k} $ is a small correction. This representation is valid everywhere except within a thin subsurface layer.
By substituting $ \thetak{k} $ and $ \thetak{k0} $
into Eqs.~(\ref{main_tet_1}) and (\ref{main_tet_2}) and linearizing these equations in $ \WI{\chi}{k} $ and $ \triangle \Lamb{e} $, we obtain
\begin{eqnarray}
\triangle_{x}(\WI{\chi}{e}+\Lamb{i}\WI{\chi}{i})&=&-(\WI{\eta}{i}\thetak{i0}^{\etak{i}{-}1}\WI{\chi}{i} \label{equat_main_lin1} \\
&&+ \Lamb{m}\Lamb{e0}\etak{e}\thetak{e0}^{\etak{e}{-}1}\WI{\chi}{e} +\Lamb{m}\triangle\Lamb{e}\thetak{e0}^{\etak{e}}), \nonumber\\
\triangle_{x}(\WI{\chi}{e}-\Lamb{m}\Lamb{i}\WI{\chi}{i})&=&-\Lamb{G}(\etak{i}\thetak{i0}^{\etak{i}{-}1}\WI{\chi}{i} \label{equat_main_lin2}\\
&&-\Lamb{e0}\etak{e}\thetak{e0}^{\etak{e}{-}1}\WI{\chi}{e} -\triangle\Lamb{e}\thetak{e0}^{\etak{e}}).\nonumber
\end{eqnarray}
The initial conditions
\begin{equation}
\WI{\chi}{k}(0)=\WI{\chi}{k}^{\prime}(0)=0 \label{init_cond}
\end{equation}
show that $ \triangle \Lamb{e} $ determines the overall scale of $ \WI{\chi}{k} $.

%%%%%%%%%%%%%%%%%%%%%%%%%%%%%%%%%%%%%%%%%%%%%%%%%%%%%%%%%%
\subsection{Irregular component in the WKB approximation}
%%%%%%%%%%%%%%%%%%%%%%%%%%%%%%%%%%%%%%%%%%%%%%%%%%%%%%%%%%

When both sides of Eq.~(\ref{equat_main_lin2}) are divided by $ \Lamb{G} $,
the Laplacian acquires the small coefficient $ 1 / \Lamb{G} $.
A similar case occurs in the Schr\"odinger wave equation, in which the kinetic term with the Laplacian contains $ \hbar^2 $.
In the semiclassical limit, $ \hbar $ is small, which makes the similarity of Eqs.~(\ref{equat_main_lin1}) and (\ref{equat_main_lin2})
to the semiclassical limit of the Schr\"odinger wave equation quite obvious. In quantum mechanics,
the Wentzel, Kramers, and Brillouin approximation is used to describe the semiclassical limit; this approach is known as the WKB method (see, e.g., \cite{LandauV3}).

Following the general scheme, we look for a solution of the form
\begin{equation}
\WI{\chi}{k}^{\mathrm{irr}}(x) = \WI{g}{k}(x) \exp\left({\sqrt{\Lamb{G}} S(x)}\right). \label{non-pertrubative}
\end{equation}
In the neighborhood of $ \Lamb{G} = \infty $, the functions $ \WI{g}{k} $ admit a series expansion in $ 1 / \sqrt{\Lamb{G}} $:
\begin{equation}
\WI{g}{k}(x) = \WI{g}{k 0}(x) + \frac{\WI{g}{k 1}(x)}{\sqrt{\Lamb{G}}} + \frac{\WI{g}{k 2}(x)}{\Lamb{G}} + \ldots . \label{gie}
\end{equation}
The exponential factor causes the point $ \Lamb{G} = \infty $ to be an essentially singular point of $ \WI{\chi}{k}^{\mathrm{irr}} (x) $.
Applying $ \triangle_{x} $ to $ \WI{\chi}{k}^{\mathrm{irr}} (x) $ yields
\begin{eqnarray}
\triangle_{x} \WI{\chi}{k}^{\mathrm{irr}} &=& \exp({\sqrt{\Lamb{G}} S })
\label{D_chi} \\
&\times& \left[\Lamb{G} \WI{g}{k} (S^{\prime} )^2  + \sqrt{\Lamb{G}} \frac{(x^2 \WI{g}{k}^2 S^{\prime})^{\prime}}{x^2 \WI{g}{k}}
+\frac{(x^2 \WI{g}{k}^{\prime})^{\prime}}{x^2}\right]. \nonumber
\end{eqnarray}
The coefficients of powers of $ \sqrt{\Lamb{G}} $ are as follows:
\begin{eqnarray}
\Lamb{G}&:&\quad \WI{g}{k0}(S^{\prime} )^2,  \quad \label{LambG1}\\
\sqrt{\Lamb{G}}&:&\quad \WI{g}{k1}(S^{\prime} )^2 + \frac{(x^2 \WI{g}{k0}^2 S^{\prime})^{\prime}}{x^2 \WI{g}{k0}},\quad \label{LambG2}\\
1&:&\quad \WI{g}{k2}(S^{\prime} )^2 + \frac{(x^2 \WI{g}{k1}^2 S^{\prime})^{\prime}}{x^2 \WI{g}{k1}}+\frac{(x^2 \WI{g}{k0}^{\prime})^{\prime}}{x^2},\quad \label{lambG3}
\end{eqnarray}
where the common exponential factor is suppressed.

The $ \Lamb{G} $ term in Eq.~(\ref{equat_main_lin1}) yields the relation
\begin{equation}
\WI{g}{e0}+\Lamb{i}\WI{g}{i0}=0. \label{g0ei}
\end{equation}
By substituting this relation into Eq.~(\ref{equat_main_lin2}), we obtain with the same accuracy
\begin{equation}
S^{\prime} = \pm \sqrt{
\frac{ \etak{i} \thetak{i0}^{\etak{i} - 1}  + \Lamb{i}\Lamb{e}^{\mathrm{reg}}\etak{e} \thetak{e0}^{\etak{e} - 1}}
 { \Lamb{i} (1 + \Lamb{m}) }
}. \label{S_prime}
\end{equation}
The exponent can be found by integrating Eq.~(\ref{S_prime}). In the model with
unit polytropic indices,
this expression gives a simple linear dependence,  $ S \sim x $.

The $ \sqrt {\Lamb{G}} $ term in Eq.~(\ref{equat_main_lin1}) leads to the relation
\begin{equation}
\WI{g}{e1}+\Lamb{i}\WI{g}{i1}=0. \label{g1ei}
\end{equation}
Using this equality in Eqs.~(\ref{equat_main_lin2}) and (\ref{S_prime}),
we find that the second differential term in Eq.~(\ref{LambG2}) vanishes. We thus obtain
\begin{equation}
x^2 \WI{g}{e0}^2 S^{\prime}=\mathrm{const}. \label{ge0}
\end{equation}
Near the center, $ \WI{g}{k0} $ behaves as $ \WI{g}{k0} \sim 1/x $.
Equation (\ref{S_prime}) leads to
\begin{equation*}
S_{\pm}(x) = \pm \int_0^{x}\sqrt{\frac{\etak{i} \thetak{i0}^{\etak{i} - 1}(x^{\prime})+
\Lamb{i} \Lamb{e}^{\mathrm{reg}} \etak{e} \thetak{e0}^{\etak{e} - 1}(x^{\prime})}{\Lamb{i} (1 + \Lamb{m})}} dx^{\prime}. \label{S}
\end{equation*}
The terms with $ S_{\pm} $ appear in the expression for $ \WI{\chi}{k0}^{\mathrm{irr}} (x ) $ in a certain combination.
The singularity at $ x = 0 $ can be eliminated by taking the difference
$ \exp \left(\sqrt{\Lamb{G}} S_{+} (x) \right) - \exp \left(\sqrt{\Lamb{G}} S_{-}(x) \right) $.
Finally,
\begin{equation}
\WI{\chi}{k0}^{\mathrm{irr}}(x) = \WI{C}{k0}\frac{\sinh\left(\sqrt{\Lamb{G}}S_{+}(x)\right)} {x\sqrt{ S_{+}^{\prime}(x)}}  + O({\Lamb{G}^{-\frac{1}{2}}}), \label{chi_1}
\end{equation}
where the $ \WI{C}{k 0} $ are arbitrary constants.

The higher-order terms of the expansion
can be found in a similar manner.

The functions $ \WI{\chi}{k0}^{\mathrm{irr}} (x) $ do not vanish at $ x = 0 $. The normalization condition $ \thetak{k}(0) = 1 $
is restored when considering a particular solution of the inhomogeneous ODE system defined by Eqs.~(\ref{equat_main_lin1}) and (\ref{equat_main_lin2}).

\subsection{Correction to the irregular component}

A particular solution to Eqs.~(\ref{equat_main_lin1}) and (\ref{equat_main_lin2}) can be found
from the original ODE system defined by Eqs.~(\ref{main_tet_1}) and (\ref{main_tet_2}).
We can construct the regular solution to this system, starting from the LEC solution,
in the form of a series expansion in $ 1 / \Lamb{G}$.
The functions $ \thetak{k0} (x) $ and $ \Lamb{e}^{\mathrm{reg}} $ that are obtained in this way
depend on the parameters $ \Lamb{i} $, $ \Lamb{m} $, and $ \Lamb{G}$.

Suppose that we are looking for the regular solution to Eqs. (\ref{main_tet_1}) and (\ref{main_tet_2})
with the initial conditions $ \thetak{k} (0) \neq 1 $ and $ \thetak{k}^{\prime} (0) = 0 $
for $|\thetak{k} (0) - 1| \ll 1 $.
Let $ \WI{\widehat{\theta}}{k0} (x) $ denote the corresponding solutions,
which depend on the parameters $ \Lamb{i} $, $ \Lamb{m} $, and $ \Lamb{G} $.
The same parameters determine $\Lamb{e} \equiv \Lamb{e}^{\mathrm{reg}\prime}$.

It is not difficult to see that by rescaling $x$ and $ \WI{\widehat\theta}{k0}(x)$,
the problem can be reduced to the initial problem with the standard initial conditions, the rescaled variable $x$,
\[
\widehat{x}=x\sqrt{\frac{\thetak{i}^{\etak{i}}(0)}{\thetak{e}(0)}},
\]
and the new parameters $\Lamb{e}$ and $\Lamb{i}$:
\begin{equation*}
%\widehat{x}=x\sqrt{\frac{\theta_{i}^{\eta_i}(0)}{\theta_{e}(0)}},\quad
\WI{\widehat{\Lambda}}{e}=\Lamb{e}^{\mathrm{reg}\prime}\frac{\thetak{e}^{\etak{e}}(0)}{\thetak{i}^{\etak{i}}(0)},\quad
\WI{\widehat{\Lambda}}{i}=\Lamb{i}\frac{\thetak{i}(0)}{\thetak{e}(0)}.
\end{equation*}
By contrast, $ \Lamb{m} $ is not modified. Obviously, $ \WI{\widehat{\Lambda}}{e}$ is a known function of $ \WI{\widehat{\Lambda}}{i} $,
namely, $ \WI{\widehat{\Lambda}}{e} = \Lamb{e}^{\mathrm{reg}} (\WI{\widehat{\Lambda}}{i}) $; thus,
\[
\Lamb{e}^{\mathrm{reg}\prime}(\Lamb{i}) = \frac{\thetak{i}^{\etak{i}}(0)}{\thetak{e}^{\etak{e}}(0)} \Lamb{e}^{\mathrm{reg}} \Big(\Lamb{i}\frac{\thetak{i}(0)}{\thetak{e}(0)}\Big).
\]
Under the assumption of the smallness of $ \triangle \thetak{k} = \thetak{k} (0) - 1$, we obtain
\begin{eqnarray*}
\triangle\Lamb{e} = \Lamb{e}^{\mathrm{reg}\prime} - \Lamb{e}^{\mathrm{reg}} %\nonumber \\
&=&\triangle\thetak{i}\left(\etak{i}\Lamb{e}^{\mathrm{reg}}+
\Lamb{i}\frac{\partial\Lamb{e}^{\mathrm{reg}}}{\partial\Lamb{i}}\right) \nonumber \\
&-&\triangle\thetak{e}\left(\etak{e}\Lamb{e}^{\mathrm{reg}}+
\Lamb{i}\frac{\partial\Lamb{e}^{\mathrm{reg}}}{\partial\Lamb{i}}\right).\label{new_Lambda}
\end{eqnarray*}
The solution to Eqs.~(\ref{main_tet_1}) and (\ref{main_tet_2}) with the non-standard boundary conditions can be written as follows:
\begin{equation}
\WI{\widehat{\theta}}{k0}(x)=\thetak{k}(0)\thetak{k0}\!\left(x\sqrt{\frac{\thetak{i}^{\etak{i}}(0)}{\thetak{e}(0)}},
\Lamb{i}\frac{\thetak{i}(0)}{\thetak{e}(0)}\right). \label{phi}
\end{equation}
The function
\begin{equation}
\WI{\chi}{k0} = \WI{\widehat{\theta}}{k0}-\thetak{k0} %,\quad\mathrm{k=(i,e)},
\end{equation}
satisfies the linearized equations (\ref{equat_main_lin1}) and (\ref{equat_main_lin2}) for $ \triangle \Lamb{e} = \Lamb{e}^{\mathrm{reg}\prime} - \Lamb{e}^{\mathrm{reg}}$ and the initial conditions
$ \WI{\chi}{k0}(0) = \triangle \thetak{k} $.
For small $\triangle \thetak{k} $, we can write
\begin{eqnarray}
\WI{\chi}{k0} (x) &=& \thetak{k0}(x)\triangle\thetak{k} + x \frac{\partial\thetak{k0}(x)}{\partial x} \frac{\etak{i}\triangle\thetak{i}{-}\triangle\thetak{e}}{2} \nonumber \\
&+& \Lamb{i} \frac{\partial\thetak{k0}(x)}{\partial \Lamb{i}} (\triangle\thetak{i}{-}\triangle\thetak{e}). \label{phi_theta}
\end{eqnarray}

\subsection{General solution}

The irregular component of the general solution is given by the first term of the expansion in Eq.~(\ref{gie}).
Equation (\ref{chi_1}) is a solution to the homogeneous equation,
whereas Eq.~(\ref{phi_theta}) is a solution to the non-homogeneous equation.
The general solution is defined by Eq.~(\ref{general}) with $\WI{\chi}{k} = \WI{\chi}{k}^{\mathrm {irr}} + \WI{\chi}{k0}$.
Among these functions, $\thetak{k0}$ and $\WI{\chi}{k0}$ are regular at $\Lamb{G} = \infty$, whereas $\WI{\chi}{k}^{\mathrm {irr}}$ is singular at $\Lamb{G} = \infty$.
The initial conditions (\ref{init_cond}) lead to the relation
\begin{equation}
\WI{C}{k0}\sqrt{\Lamb{G}S_{+}^{\prime}(0))}+\triangle\thetak{k}=0, %\quad\mathrm{k=(i,e)},
\end{equation}
where the $\WI{C}{k0} $ are the constants that appear in Eq.~(\ref{chi_1}).
The derivative of $ \WI{\chi}{k0} $ at $ x = 0 $ vanishes by construction.
In the general case, $ S_{\pm} (x) = x S_{\pm}^{\prime} (0) + O (x^3) $;
consequently, the derivative $ \WI{\chi}{k}^{\mathrm{irr}} $ also vanishes automatically at $ x = 0 $.
As a consequence of Eq.~(\ref{g0ei}), the coefficients $ \WI{C}{k0} $ satisfy
\[
\WI{C}{e0} + \Lamb{i}\WI{C}{i0}=0.
\]
By expressing the unknown coefficients in terms of $ \WI{C}{i0} $, we can write
\begin{widetext}
\begin{eqnarray*}
\WI{\chi}{i}(x) &=& \frac{\WI{C}{i0}}{x\sqrt{S_{+}^{\prime}(x)}} \sinh{\left(\sqrt{\Lamb{G}} S_{+} (x)\right)}
- \WI{C}{i0}\sqrt{\Lamb{G}S_{+}^{\prime}(0)}\Bigg[\thetak{i0}(x) + x \frac{\partial\thetak{i0}(x)}{\partial x}
\frac{\etak{i}{+}\Lamb{i}}{2}+\Lamb{i} \frac{\partial\thetak{i0}(x)}{\partial \Lamb{i}} (1 {+} \Lamb{i})\Bigg], \label{fin i} \\
\WI{\chi}{e}(x) &=&-\frac{\WI{C}{i0}\Lamb{i}}{x\sqrt{S_{+}^{\prime}(x)}}\sinh{\left(\sqrt{\Lamb{G}}S_{+}(x)\right)}
+ \WI{C}{i0}\sqrt{\Lamb{G}S_{+}^{\prime}(0)}\Bigg[\Lamb{i}\thetak{e0}(x) -
 \frac{\partial\thetak{e0}(x)}{\partial x} \frac{\etak{i}{+}\Lamb{i}}{2}-\Lamb{i} \frac{\partial\thetak{e0}(x)}{\partial \Lamb{i}} (1{+}\Lamb{i})\Bigg].  \label{fin e}
\end{eqnarray*}
\end{widetext}
In each of these expressions, the first term increases exponentially with $ x $, dominates the second (regular) term for $ x \gtrsim 1/\sqrt{\Lamb{G}} $,
and becomes comparable to the regular component $\thetak{k0}$ near the surface in the layer $ | x - \WI{x}{b} | \sim \thetak{eb} $,
 %or $\sim \Lambda_{G}^{- 1/(1 + \eta_i)}$,
where perturbation theory in $ \WI{\chi}{k} $ is no longer applicable.
The constant $ \WI{C}{i0} $ is exponentially small.
This feature has been noted earlier in the analysis of the exactly solvable model of Sect.~IV.
The exponential smallness of the irregular solution in the inner layers of the star is obviously a common property.
The functions $ \WI{\chi}{i} (x) $ and $ \WI{\chi}{e} (x) $ differ in sign, and the electron component is suppressed by approximately $ \Lamb{i} $.

The matching of $\thetak{k}$ and $\thetak{k}^{\prime}$ with the electrosphere solution requires
\[
\WI{\chi}{e}(\WI{x}{b}) \sim \thetak{eb}, \quad {\WI{\chi}{e}^{\prime}(\WI{x}{b})} \sim 1.
\]
The first of the equations allows to fix $\WI{C}{i0}$.
For $\thetak{i}(\WI{x}{b}) = 0$, Eq.~(\ref{S_prime}) leads to $S_{+}^{\prime}(\WI{x}{b}) \sim \thetak{eb}^{(\etak{e} - 1)/2}$;
the second equation then follows automatically.
One can verify that for $\WI{\eta}{e} = \WI{\eta}{i} = 1$, $ \WI{\chi}{k}^{\mathrm{irr}} $
is analytic in $\Lamb{G} \in \mathbb{C}^1 \setminus \infty$, with $\Lamb{G} = \infty$ being a unique singular point of $ \WI{\chi}{k}^{\mathrm{irr}} $,
in agreement with the analysis of Sect.~IV.B.

Standard numerical methods for solving ODE systems do not allow the separation of the regular solution from the irregular one
due to the exponential smallness of the latter. A numerical solution to Eqs.~(\ref{main_tet_1}) and (\ref{main_tet_2}) for $ x \lesssim 1 $
will inevitably contain an admixture of the irregular component, which is interpreted by the program as a machine zero.
In the subsurface layer $ | x - \WI{x}{b} | \sim \thetak{eb} $, due to the exponential growth of $ \WI{\chi}{k}^{\mathrm{irr}} (x) $,
this machine zero is converted into a finite but random value.
The stellar properties that are sensitive to $ \triangle \Lamb{e}$ cannot be determined using numerical methods.
A change in the integration, e.g., of the grid step, is thus expected to result in uncontrolled changes in the stellar charge, the thickness of the electro- or ionosphere, and other quantities that depend on $ \triangle \Lamb{e} $. In view of these remarks,
the numerical solutions of Refs.~\cite{Rotondo2011,Belvedere2012,Belvedere2014} are likely to be unstable.

At the same time, as the example of Eqs.~(\ref{charge 1}) and (\ref{charge 2}) shows,
relationships which do not depend on $ \triangle \Lamb{e} $ are numerically stable.
Each integration with a varying grid step
leads to random values of $ \WI{q}{s} $ and $ \thetak{eb} $, but all points on the plane $ (\WI{q}{s}, \thetak{eb}) $ must lie on a smooth curve.
In the exactly solvable model of Sect.~IV, the quantities $ \WI{q}{s} $ and $ \thetak{eb} $ are related by a linear dependence.

%%%%%%%%%%%%%%%%%%%%%%%%%%%%%%%%%%%%%%%%%%%%%%%%%%%%%%%%%%% %%%%%%%%%%%%%%%%%%%%%%%%%%
%%%%%%%%%%%%%%%%%%%%%%%%%%%%%%%%%%%%%%%%%%%%%%%%%%%%%%%%%%% %%%%%%%%%%%%%%%%%%%%%%%%%%
\section {Conclusions}
%%%%%%%%%%%%%%%%%%%%%%%%%%%%%%%%%%%%%%%%%%%%%%%%%%%%%%%%%%% %%%%%%%%%%%%%%%%%%%%%%%%%%
%%%%%%%%%%%%%%%%%%%%%%%%%%%%%%%%%%%%%%%%%%%%%%%%%%%%%%%%%%% %%%%%%%%%%%%%%%%%%%%%%%%%%
\setcounter {equation} {0}

In this paper, we considered the unconstrained HE equations for stars in the absence of local electroneutrality.
The condition of local electroneutrality is usually imposed on the EoS phenomenologically as an additional constraint (LEC).
In our approach, the electrostatic interactions are considered \textit{ab initio};
as a result, the LEC is satisfied approximately, with the uncompensated charge density being $ 1/\WI{\lambda}{G} \sim 10^{-36} $
of the particle number density.
The smallness of the parameter $ 1/\WI{\lambda}{G} $ makes it possible to use the LEC to describe with a high accuracy properties that are not related to stellar electrostatics.

We found that the mass-radius relationship and the condition for gravitational stability for charged stars
are not modified to an accuracy of $ \sim \WI{\lambda}{G}^{-1/(1 + \etak{e})}$.
The admissible states of charged stars are determined by the condition that
the electro- and ionospheres do not extend to infinity.
This  requirement provides lower and upper bounds on the total stellar charge of $-0.1$ and $150$ C per solar mass, respectively.

The main system of unconstrained HE equations, Eqs.~(II.12) and (II.13), belongs to the class of singularly perturbed ODEs.
In such a system, the perturbation parameter (in our case, $ 1/\WI{\lambda}{G} $)
appears as a coefficient of the higher derivative and thus plays the main role in specifying the nature of the problem.
The use of conventional (regular) methods for solving ODEs via series expansion in a small parameter
leads to the loss of certain solutions in this case.
In the limit of $\WI{\lambda}{G} \to \infty$, Eq.~(II.13) becomes a constraint on the particle concentrations,
which we identify as an LEC.

The Poincar\'e theorem on analyticity and Dyson's argument indicate that the general solution to Eqs.~(II.12) - (II.13)
is singular for $\WI{\lambda}{G} = \infty$.
A particular, regular solution can exist provided that one of the parameters, e.g., $\Lamb{e}$,
is replaced by a function of $\WI{\lambda}{G}$.
Such a solution can be constructed as a formal power series in $G$ ($\sim 1/\WI{\lambda}{G}$) starting from a locally neutral solution.
Only these two types of solutions, namely, regular ones and locally neutral ones, have been discussed in the literature to date.

In this paper, the general solution to the HE problem was first constructed in NGT.

In Sect. IV, we presented an exactly solvable model of a two-component mixture of ions and electrons with unit polytropic indices.
The general solution to the unconstrained HE equations appears to be irregular in the neighborhood
of $ \WI{\lambda}{G} = \infty $. The irregular component is exponentially small in the inner layers of the star.

In the general case, the WKB method can be used to construct the irregular component.
Its magnitude is comparable to that of the regular component in a subsurface layer of thickness $ \sim \WI{\lambda}{G}^{-1/(1 + \etak{e})} $,
where these components interact non-linearly and determine the structure of the electro- or ionosphere.
The properties of these envelopes are {exponentially} sensitive to variations in the charge density in the central regions of the star.

In general, stars are characterized by their mass, radius and charge.
Nuclear reactions in the interior of a star cause variations in the parameters $\Lamb{e}$, $\Lamb{i}$, and $\Lamb{m}$
and thereby affect the bulk stellar charge,
which, due to the conservation of electric charge, leads to a charge exchange between the envelope and the inner region.

The structure of the general solution to the unconstrained HE equations
can be described as follows:

Starting from a locally neutral solution, one can construct a regular solution  ($\thetak{k0}$ in Sect.~V) in the form of a series in powers of $G$.
On the basis of this solution, an irregular component can be constructed using the WKB method ($\WI{\chi}{k}^{\mathrm{irr}}$ in Sect.~V).
This component is singular for $\WI{\lambda}{G} = \infty$.
A correction to this irregular component is needed to satisfy the initial conditions ($\WI{\chi}{k0}$ in Sect.~V).
The sum of these three functions yields the general solution, excluding only a thin subsurface layer of thickness $ \sim \WI{\lambda}{G}^{-1/(1 + \etak{e})} $,
where the magnitude of the irregular component becomes comparable to that of the regular component and the WKB method is no longer applicable.
Numerical methods should then be used to match, within the subsurface layer, the general solution
with the known solution for the electro- or ionosphere.

Extension of the formalism developed here to GRT
is left as an open problem.
An important direction of study is generalization to
three or more component fluids in chemical equilibrium.
Yet another interesting application is to find the general solution of the unconstrained HE equations for strange stars.

\vspace{12pt}
The authors are grateful to S.~I.~Blinnikov and I.~L.~Iosilevsky for stimulating discussions.
M.I.K. wishes to acknowledge M.~D.~Malykh for discussion of analytic properties of the solutions.
The work is supported in part by the RFBR Grants Nos. 16-02-00228 and 18-02-00733.

\appendix

\section{Incompressible two-fluid model} \label{appen1}
\renewcommand{\theequation}{A.\arabic{equation}}
\setcounter{equation}{0}

As a second example of an exactly solvable, albeit somewhat exotic, model,
we consider the case of $ \WI{\eta}{i} = \WI{\eta}{e} = 0 $, which corresponds to an incompressible fluid
($ n = n_0 \theta^{\eta} \sim$ const). %,
Although the speed of sound in such a substance is infinite,
this case is of methodological interest with regard to the restrictions on the parameter $ \Lamb{e} $ in the regular solution,
especially as $ S_{\pm} (x) \to 0 $ for $ \WI{\eta}{i} = \WI{\eta}{e} = 0 $.
For an incompressible fluid, Eqs.~(\ref{main_tet_1}) and (\ref{main_tet_2}) become
\begin{eqnarray}
\triangle_{x}(\thetak{e}+\Lamb{i}\thetak{i})&=&-(1{+}\Lamb{m}\Lamb{e}),\label{equat_01}\\
\triangle_{x}(\thetak{e}-\Lamb{m}\Lamb{i}\thetak{i})&=&-\Lamb{G}(1{-}\Lamb{e}).\label{equat_02}
\end{eqnarray}
Their solutions for the initial conditions $ \thetak{k} (0) = 1 $ and $ \thetak{k}^{\prime}(0) = 0 $ have the form
\begin{eqnarray}
\thetak{i}&=&1-\frac{(1{+}\Lamb{m}\Lamb{e})-\Lamb{G}(1{-}\Lamb{e})}{6\Lamb{i}(1{+}\Lamb{m})} x^2, \label{theti} \\
\thetak{e}&=&1-\frac{\Lamb{m}(1{+}\Lamb{m}\Lamb{e})+\Lamb{G}(1{-}\Lamb{e})}{6(1{+}\Lamb{m})} x^2. \label{thete}
\end{eqnarray}
In principle, these solutions describe a charged stellar configuration free from constraints on $ \Lamb{e} $.
This freedom is due to the irregular nature of the solution for $ \Lamb{G} = \infty $.
The requirement of regularity leads to the LEC $ \Lamb{e} = 1 $, given that $ \Lamb{e}$ is a parameter.
If $ \Lamb{e}$ is considered to be a function of $ \Lamb{G}$, then $ \Lamb{e} = 1 + O(1/\Lamb{G})$.
The solutions (\ref{theti}) and (\ref{thete}) are valid in the inner region of the star.
The outer region requires special consideration.

According to Eq.~(\ref{theti}), the ion concentration vanishes for
\begin{equation}
\WI{x}{i}=\sqrt{\frac{6\Lamb{i}(1{+}\Lamb{m})}{(1{+}\Lamb{m}\Lamb{e})-\Lamb{G}(1{-}\Lamb{e})}}.\label{x_i}
\end{equation}
If we require the electron concentration to vanish at the same point, $ \WI{x}{e} = \WI{x}{i} $,
then we obtain $ \Lamb{e} = \Lamb{e}^{\mathrm{reg}}$, where $\Lamb{e}^{\mathrm{reg}}$ is given by Eq.~(\ref{ni_eq_ne_Lambe}).
In this case, the star still has a charge. Suppose now that we do not require the concentrations to vanish simultaneously.
Again, let us consider the outer part of the star, the electrosphere, where the ion concentration is zero.
The equation for $\thetak{e}$ reduces to
\begin{equation}
(1 + \Lamb{m})\triangle_{x}(\thetak{e})=\Lamb{e}(\Lamb{G}{-}\Lamb{m}^2). \label{equat_electron}
\end{equation}
At the boundary $ x = \WI{x}{i} $, the function $ \thetak{e} $ and its first derivative are continuous.
The solution to Eq.~(\ref{equat_electron}) takes the form
\begin{equation*}
\thetak{ea}=1-\frac{\WI{x}{i}^2(\Lamb{G}{+}\Lamb{m})}{2(1{+}\Lamb{m})}\left(1-\frac{2\WI{x}{i}}{3x}\right)
+\frac{\Lamb{e}(\Lamb{G}{-}\Lamb{m}^2)}{6(1{+}\Lamb{m})}x^2. \label{thete_atm}
\end{equation*}
For an electrically neutral star, the boundary $ x = \WI{x}{e} $, where $ \thetak{ea} (\WI{x}{e}) = 0 $,
determines the total charge. The bulk charge is proportional to
(cf. Eq.~(\ref{Charge}))
\begin{equation}
\WI{Q}{s} \sim -x^2(\thetak{e}'{-}\Lamb{m}\Lamb{i}\thetak{i}')\Big|^{\WI{x}{i}}_0,\label{Q_s}
\end{equation}
whereas the charge of the envelope is proportional to
\begin{equation}
\WI{Q}{e} \sim -\frac{\Lamb{G}(1{+}\Lamb{m})}{\Lamb{G}{-}\Lamb{m}^2}x^2\WI{\theta}{ea}^{\prime}\Big|^{\WI{x}{e}}_{\WI{x}{i}}.\label{Q_a}
\end{equation}
The condition $ \WI{Q}{s} + \WI{Q}{e} = 0$ leads to the relation
\begin{equation}
\WI{x}{e}=\WI{x}{i}/{\sqrt[3]{\Lamb{e}}}, \label{x_e}
\end{equation}
whereas the condition $ \WI{\theta}{ea} (\WI{x}{e}) = 0 $ leads to the following equation:
\begin{eqnarray}
&\;&\Lamb{e}(\Lamb{G}{+}\Lamb{m})+\Lamb{i}\sqrt[3]{\Lamb{e}}(3\Lamb{G}{+}2\Lamb{m}{-}\Lamb{m}^2) \nonumber \\
&&\;\;\;\;\;\;\;\; \;\;\;\;=
\Lamb{G}{-}1+3\Lamb{i}(\Lamb{G}{+}\Lamb{m}), \label{Lamb_e_total_0}
\end{eqnarray}
whose solution, $\Lamb{e}$, coincides with $\Lamb{e}^{\mathrm{reg}}$ to an accuracy of $1/\Lamb{G}$.

\end{document}